\documentclass[preprintnumbers, floatfix, twocolumn,
preprintnumbers, PRD, superscriptaddress,nofootinbib]{revtex4-1}
\usepackage[utf8]{inputenc}
\usepackage{graphicx}
\usepackage{epsfig}
 \usepackage{amsmath,amssymb}
\usepackage[colorlinks]{hyperref}
\usepackage[usenames,dvipsnames]{color}
\hypersetup{
     breaklinks=true,
    pdfstartview={FitH},    
    colorlinks=true,       
    linkcolor=blue,          
    citecolor=red,        
    filecolor=magenta,      
    urlcolor=blue,           
    anchorcolor=green,      
    linktocpage=true
}

\newcommand{\be}{\begin{equation}}
\newcommand{\ee}{\end{equation}}
\begin{document}
\title{Cosmology from String T-duality and zero-point length}

\author{G.~G.~Luciano}
\email{giuseppegaetano.luciano@udl.cat} 
\affiliation{Department of Chemistry, Physics and Environmental and Soil Sciences, Escola Politecninca Superior, Universidad de Lleida, Av. Jaume
II, 69, 25001 Lleida, Spain}

\author{A. Sheykhi}
\email{asheykhi@shirazu.ac.ir} \affiliation{Department of Physics,
College of Science, Shiraz University, Shiraz 71454, Iran}
\affiliation{Biruni Observatory, College of Science, Shiraz
University, Shiraz 71454, Iran}

\date{\today}

\begin{abstract}
Inspired by String T-duality and taking into account the zero-point
length correction, $l_0$, to the gravitational potential, we construct 
modified Friedmann equations by applying the first law of
thermodynamics on the apparent horizon of the
Friedmann-Robertson-Walker (FRW) Universe. The cosmological viability of 
this extended scenario is investigated by studying influences on the evolution of the early Universe and performing the cosmographic analysis. Furthermore, we explore the inflationary paradigm under the slow-roll condition. 
By testing the model against observational data, the zero-point length  
is constrained around the Planck scale, in compliance with the original assumption
from String T-duality. We also study the growth of density perturbations in the linear regime. It is shown that the zero-point length stands out as an alternative
characterization for the broken-power-law spectrum, which 
provides a better fitting for the experimental measurements comparing to the simple power-law potential. Finally, we focus on implications for the primordial gravitational waves (PGWs) spectrum. Should the zero-point length be running over energy scales, as is the case for all parameters and coupling constants in quantum gravity under renormalization group considerations, deviations from General Relativity (GR) might be potentially tested through upcoming PGW observatories. 
\end{abstract}
 \maketitle

\section{Introduction \label{Intr}}

Nowadays, it is a general belief that laws of gravity may be
regarded as a manifestation of laws of thermodynamics for the
large scale spacetime systems. This idea has been well established
in the past three decades, where it has been confirmed that the
gravitational field equations can be constructed from the first
law of thermodynamics on the boundary of
spacetime~\cite{Jac,Pad1,Pad2}. This approach is also reversible,
namely one can start from gravitational field equations governing
dynamics of the system in the bulk and show that it can be
rewritten in the form of the first law of thermodynamics on the
boundary of the system~\cite{Pad3}. The deep connection between
thermodynamics and gravity has been explored in different setups. In particular, in the context of FRW cosmology, it has been
shown that the Friedmann equations describing the evolution of the
Universe can be rewritten as the first law of thermodynamics on
the apparent horizon and vice
versa~\cite{CaiKim,Cai2,wang1,SheyWC,SheyWC2,SheyCQ}.

Although GR has successfully passed a number of experimental and
observational tests, it still suffers from some problems such as
curvature singularity at the center of black holes or initial
singularity in the standard cosmology. Removing these
singularities, using the first principles and fundamental theories
such as string theory and quantum gravity, is one of the main
challenges of modern theoretical physics. To relax singularity in
the black hole solutions, regular black holes were proposed by
combining the non-linear electrodynamics with Einstein-Hilbert
Lagrangian~\cite{RegBH1,RegBH2}. An attempt towards constructing
singularity free black holes was performed by averaging
noncommutative coordinates on suitable coherent states in string
theory~\cite{Nico1,Nico2,Nico3}. It was argued that the effects of
noncommutative is equivalent to a nonlocal deformation of the
Einstein-Hilbert action~\cite{Nico4,Mod}.

Another approach to deal with nonsingular black holes comes from
the concept of T-duality in string theory. It was argued that the
momentum space propagator induced by the path integral duality,
for a particle with mass $m_0$, is given by~\cite{Padm,Sma,Spa,Fon}
\begin{eqnarray}
G(p)=-\frac{l_0 }{\sqrt{p^2+m_0^{2}}} K_{1}\left(l_0
\sqrt{p^2+m_0^{2}}\right),
\end{eqnarray}
where $p^2=(\hbar k)^2$ is now the squared momentum, $l_0$
represents the zero-point length of spacetime and $K_{1}(x)$ is a
modified Bessel function of the second kind. For the massless
particles the above propagator can be written as ($\hbar=1$)~\cite{td1}
\begin{eqnarray}
G(k)=-\frac{l_0 }{\sqrt{k^2}} K_{1}\left(l_0 \sqrt{k^2}\right).
\end{eqnarray}
In the limiting case where $(l_0 k^2)\rightarrow 0$, one find
$G(k)=-k^{-2}$ for the massless propagator.

The static Newtonian
potential corresponding to $G(k)$ at distance $r$ is obtained as~\cite{td1}
\begin{eqnarray}
\label{V}
V(r)=-\frac{M }{\sqrt{r^2+l_0^2}},
\end{eqnarray}
where $M$ is the mass of the source of the potential. It has been
shown that this finite potential changes the metric around an
electrically neutral black hole and leads to a regular spacetime
at $r=0$~\cite{td1}. Besides, it modifies the Hawking temperature,
the entropy associated with the horizon, and hence thermodynamics
of black holes~\cite{td1}.

Following the entropic force scenario proposed by
Verlinde~\cite{Ver}, the effects of zero-point length on the
Newton's law of gravity were explored in~\cite{KS}. Nevertheless,
since cosmology can explore physics at higher energy scales, it
seems natural that the implications of the zero-point length could
be more likely tested in this context. On the cosmological scales,
Eq.~\eqref{V} along with Verlinde's conjecture lead to correction
terms to the Newtonian cosmology, while in the relativistic regime
they give rise to modified Friedmann equations in a FRW
background~\cite{KS}. This procedure also leads to a modified
expression for the entropy, due to the zero-point length,
associated with the horizon, which is useful in studying
thermodynamics of the Universe. In passing, we mention that the
hypothesis of modified horizon entropy is widespread across many
extended theories of gravity, see for
instance~\cite{ME1,ME2,ME3,ME4,ME6,ME7,ME8} and references
therein.

Using thermodynamics-gravity conjecture and applying the first law
of thermodynamics on the apparent horizon, in this work we derive
the modified cosmological field equations in the presence of
zero-point length from string T-duality. Research in this
direction has been active in the last three
decades~\cite{Q0,Q1,Q2,Q3}, focusing primarily on the development
of a minimal-length cosmology in the framework of the generalized
uncertainty
principle~\cite{Lit1,Lit2,Lit3,Lit4,Lit5,Lit6,Lit7,Lit8,Lit9,Lit10}.
Compared to the previous efforts, our analysis provides an
advancement in understanding the semiclassical regime of quantum
cosmology. Indeed, on the one hand we offer an alternative
perspective by embedding a minimal length from a novel theoretical
standpoint. On the other hand, we explore phenomenological
consequences of the obtained modified field equations on a wide
range of energy scales, including implications for the dynamics of
the Universe in the radiation and matter dominated eras, the early
inflation, the growth of perturbations and gravitational waves. To
the best of our knowledge, such a dedicated analysis has not yet
been considered in literature. On top of that, we extract a
constraint on the zero-point length. The result aligns with the
original assumption that this cutoff scale is identifiable with
the Planck length~\cite{Padm}, as well as with the recent estimate
of the zero-point length in the context of quantum corrected black
holes~\cite{td1}.

The work is structured as follows. In the next section we develop
a modified cosmological scenario based on the zero-point length
and study its implications for the evolution of the Universe.
Section~\ref{inflation} is devoted to analyze of inflation, while
in Sec.~\ref{pert} we focus on the growth of perturbations and
structure formation. Effects on the spectrum of primordial
gravitational waves are explored in Sec.~\ref{GW}. The discussion
of closing remarks and perspectives is reserved for
Sec.~\ref{Clos}. Unless otherwise stated, we shall use Planck
units $\hslash=c=G=k_B=1$.
\section{Modified Friedman Equations through zero-point length \label{Fried}}
According to T-duality the zero-point length affects the
gravitational potential~\cite{td1}. Thus, the gravitational
potential of a spherical mass distribution $M$ at distance $r=R$
is modified as~\cite{td2,td3}
\begin{eqnarray}
\phi(r)=-\frac{M }{\sqrt{r^2+l_0^2}}|_{r=R},
\end{eqnarray}
where $l_0$ is the zero-point length and its value is expected to
be of Planck length~\cite{td1}. Thus the Newton's law of gravity
for a mass point $m$ get modified as

\begin{equation}\label{F5}
\vec{F}=-m \vec{\nabla}
\phi(r)|_{r=R}=-\frac{Mm}{R^2}\left[1+\frac{l_0^2}{R^2}\right]^{-3/2}
\hat{r}.
\end{equation}
Using the entropic force scenario, this leads to correction to the
entropy associated with the horizon as~\cite{KS}
\begin{eqnarray}\label{ds}
1+\left(\frac{1}{2 \pi
R}\right)\frac{d{s}}{dR}=\left(1+\frac{l_0^2}{R^2}\right)^{-3/2}
\end{eqnarray}
 where $s$ is the correction term to the total entropy of the
 horizon,
\begin{eqnarray} \label{S}
 S_{h}=\frac{A}{4}+s(A).
\end{eqnarray}
and $A=4\pi R^2$ is the area of the horizon enveloped the volume
$V=4\pi R^3/3$. Combining Eqs. (\ref{ds}) and (\ref{S}), we arrive
at
\begin{eqnarray}\label{dS}
dS_h=2\pi R \left(1+\frac{l_0^2}{R^2}\right)^{-3/2} dR\,.
\end{eqnarray}

Before moving onto the cosmological analysis, 
it is noteworthy that deviations from the area-law scaling
are very common in quantum gravity, e.g. in string theory~\cite{Log1}, 
loop quantum gravity~\cite{Log2}, AdS/CFT correspondence~\cite{Carlip}, generalized uncertainty principle~\cite{Log3}, etc. Additionally, 
entropic corrections appear from entanglement entropy calculations in four spacetime dimensions at UV scale~\cite{UV1,UV2}. Therefore, 
despite the specific String T-duality framework we are considering here, 
our next arguments are expected to have more general validity  
in the context of quantum gravity. 

Let us assume our Universe is described by the line element
\begin{equation}
ds^2={h}_{\mu \nu}dx^{\mu} dx^{\nu}+R^2(d\theta^2+\sin^2\theta
d\phi^2),
\end{equation}
where $R=a(t)r$, $x^0=t, x^1=r$, and $h_{\mu \nu}$=diag $(-1,
a^2/(1-kr^2))$ stands for the two dimensional metric. $k = -1,0,
1$ represents the curvature of the Universe and $a$ the (time-dependent) 
scale factor. The apparent horizon
is the suitable boundary of the Universe from thermodynamic
viewpoint, which has radius~\cite{SheyLog}
\begin{equation}
\label{radius}
 R=\frac{1}{\sqrt{H^2+k/a^2}}.
\end{equation}
The associated temperature with the horizon is given by~\cite{Cai2,Sheyem}
\begin{equation}\label{Th}
T_h=-\frac{1}{2 \pi R}\left(1-\frac{\dot{R}}{2H R}\right),
\end{equation}
where the overdot denotes derivative with respect to the cosmic time $t$.
For an adiabatic expansion, one may assume 
$\dot{R}\ll2HR$ in the infinitesimal interval $dt$, 
which physically implies that the apparent horizon radius
is kept nearly fixed~\cite{CaiKim}.

The total energy momentum tensor is assumed in the form of the perfect
fluid
\begin{equation}
\label{T1}
T_{\mu\nu}=(\rho+p)u_{\mu}u_{\nu}+pg_{\mu\nu},
\end{equation}
where $\rho$ and $p$ are the energy density and pressure,
respectively. The conservation of energy for the metric leads to
\begin{equation}
\label{Cont}
\dot{\rho}+3H(\rho+p)=0,
\end{equation}
where $H=\dot{a}/a$ is the Hubble parameter. The work density due
to the volume change of the Universe is~\cite{Hay2}
\begin{equation}
\label{Work2}
W=-\frac{1}{2} T^{\mu\nu}h_{\mu\nu}=\frac{1}{2}(\rho-p).
\end{equation}
Taking differential form of the total energy $E=\rho (4\pi
R^3/3)$ and using the conservation equation (\ref{Cont}), we find
\begin{equation}
\label{dE2}
 dE=4\pi\tilde
 {r}_{A}^{2}\rho d\tilde {r}_{A}-4\pi H \tilde{r}_{A}^{3}(\rho+p) dt.
\end{equation}
Inserting Eqs. (\ref{dS}), (\ref{Th}), (\ref{Work2}), (\ref{dE2})
in  the first law of thermodynamics on the apparent horizon, $dE =
T_h dS_h + WdV$, and using the definition (\ref{Th}) for the
temperature, after some calculations, one gets
\begin{equation} \label{Fried1}
4\pi H R^3 (\rho+p)dt=\left(1+\frac{l_0^2}{R^2}\right)^{-3/2} dR.
\end{equation}
Using the continuity equation (\ref{Cont}), we arrive at
\begin{equation} \label{Fried2}
-\frac{2}{R^3}\left(1+\frac{l_0^2}{R^2}\right)^{-3/2} dR=
 \frac{8\pi}{3}d\rho.
\end{equation}
After integrating, we have
\begin{equation} \label{Fried3}
-\frac{2}{l_0^2}\left(1+\frac{l_0^2}{R^2}\right)^{-1/2}+C=\frac{8\pi}{3}\rho,
\end{equation}
where $C$ is the constant of integration. Since ${l_0^2}/{R^2}$ is expected to be small, we can expand the LHS of the above equations. We find
\begin{equation} 
\label{Frie4}
-\frac{2}{l_0^2}\left[ 1-
\frac{l_0^2}{2R^2}+\frac{3}{8}\frac{l_0^4}{R^4}+\mathcal{O}\left(\frac{l_0}{R}\right)^6\right]+C=\frac{8\pi}{3}\rho.
\end{equation}
Clearly, in the limiting case where $l_0^2\rightarrow 0$, we must
restore the standard Friedmann equation. Thus, we set
\begin{equation}
C\equiv \frac{2}{l_0^2}-\frac{\Lambda}{3},
\end{equation}
where $\Lambda$ is the cosmological constant. 
The final form of the modified Friedmann equation is obtained as
\begin{equation} \label{Frie5}
H^2+\frac{k}{a^2}-\alpha
\left(H^2+\frac{k}{a^2}\right)^2+\mathcal{O}(\alpha^2)=\frac{8\pi}{3}\rho+\frac{\Lambda}{3},
\end{equation}
where $\alpha=3l_0^2/4$, and we have used relation (\ref{radius}).
In a flat Universe ($k=0$), the above Friedmann equation reads
\begin{equation} \label{Frie6}
H^2-\alpha H^4=\frac{8\pi}{3}\rho+\frac{\Lambda}{3}.
\end{equation}

As we can see, the use of the modified entropy~\eqref{S} results
to a modified Friedmann equation, with an extra term comparing to
GR. In passing, we notice that such
correction could be re-interpreted as an effective
\emph{dark energy} sector by recasting Eq.~\eqref{Frie6} as
\begin{equation}
\label{de}
H^2=\frac{8\pi}{3}\left(\rho+\rho_{DE}\right),
\end{equation}
where 
\be
\rho_{DE}=\frac{3}{8\pi}\left(\frac{\Lambda}{3}+\alpha H^4\right). 
\ee
Similarly, the modified second Friedmann equation can be derived
by differentiating Eq.~\eqref{de} with respect to $t$ and using Eq.~\eqref{Cont}.
After some algebra, we obtain
\be
\label{SMF}
\dot H=-4\pi\left[\rho+p+\rho_{DE}+p_{DE}\right],
\ee 
where we have defined the pressure of the effective dark energy sector as
\be
\label{pressDE}
p_{DE}=-\frac{1}{8\pi}\left[\Lambda+\alpha H^2\left(4\dot H+3H^2\right)\right].
\ee
Hence, the effective equation of state becomes
\be
w_{DE}=\frac{p_{DE}}{\rho_{DE}}=-1-\frac{4\alpha H^2 \dot H}{3\alpha H^4+\Lambda}=-1-\frac{4\alpha H^2\dot H}{\Lambda}+\mathcal{O}(\alpha^2).
\ee

We would like to notice, however, that in the forthcoming analysis 
we do not conceptualize the zero-point length correction as an effective dark energy contribute.
Instead, we adhere to the original interpretation of T-duality and consider 
the additional term as due to a modified description of gravity. 

As expected, in the case $\alpha=0$, the modified Friedmann equations~\eqref{de} and~\eqref{SMF} reduce to the $\Lambda \textrm{CDM}$ paradigm 
\begin{eqnarray}
H^2&=&\frac{8\pi}{3}\rho+\frac{\Lambda}{3}\,,
\\[2mm]
\dot H&=&-4\pi\left(\rho+p\right). 
\end{eqnarray}

\subsection{Cosmological solutions}
Now we study the cosmological implications of the zero-point length 
scenario. We shall consider the interesting case where the perfect fluid 
includes dust matter ($p_m\approx0$) and radiation. 
From the continuity equation, these components components evolve according to
\begin{eqnarray}
\label{rm}
\rho_m&=&\rho_{m,0}a^{-3}\,,\\[2mm]
\rho_r&=&\rho_{r,0}a^{-4}\,,
\label{rr}
\end{eqnarray}
where the subscript ``0'' denotes the values at present time and we set $a_0=1$. 

Dividing both side of Eq.~(\ref{Frie6})
by $H_0^2$, we find
\begin{equation} \label{Frim1}
-\beta
\left(\frac{H}{H_0}\right)^4+\left(\frac{H}{H_0}\right)^2-\Omega_{m,0}
a^{-3}-\Omega_{r,0}
a^{-4}-\Omega_{\Lambda,0}=0\,,
\end{equation}
where 
\be
\label{beta}
\beta\equiv\alpha H_0 ^2=\frac{3}{4}l_0^2 H_0 ^2\ll1\,,
\ee 
and we have
introduced the dimensionless density parameters
\begin{eqnarray}
\Omega_i&=&\frac{8\pi}{3H^2}\rho_i,\quad (i=m,r)\,,\\[2mm]
\Omega_{\Lambda}&=&\frac{\Lambda}{3H^2}\,.
\end{eqnarray}

To convert $\beta$ in terms of the dimensional zero-point length,
it proves convenient to restore for a while the gravitational constant
$G=\ell_p^2=6.7\times10^{-39}\,\mathrm{GeV}^{-2}$, where $\ell_p$ is the Planck length. For $H_0\simeq10^{-42}\,\mathrm{GeV}$, we find $l_0\simeq1.4\times10^{61}\sqrt{\beta}\ell_p$. Coming back to our original 
units, we have 
\be
\label{convers}
l_0\simeq1.4\times10^{61}\sqrt{\beta}\,.
\ee 
Therefore, $l_0\sim\mathcal{O}(1)$ corresponds to $\beta\sim\mathcal{O}(10^{-122})$. 

Now, solving Eq.~\eqref{Frim1}, we are led to
\begin{equation} \label{Frim2}
\left(\frac{H}{H_0}\right)^2=\frac{1}{2\beta}\left\{1-
\sqrt{1-4\beta\left[ \Omega_{m,0} a^{-3}+\Omega_{r,0} a^{-4}+\Omega_{\Lambda,0}\right]}\right\}.
\end{equation}
Since $\beta$ is expected to be small, we can expand the RHS to the linear order. The result is
\begin{eqnarray}
\label{Frim5}
&&\frac{H^2}{H_0^2}=\left(\Omega_{m,0} a^{-3}+\Omega_{r,0} a^{-4}+\Omega_{\Lambda,0}\right)\\[2mm]
\nonumber
&&\times\left[1+\beta\left(\Omega_{m,0} a^{-3}+\Omega_{r,0} a^{-4}+\Omega_{\Lambda,0}\right)\right]+\mathcal{O}(\beta^2)\,.
\end{eqnarray}

If we define, as usual, the redshift parameter as
\be
\frac{1}{a}=1+z\,,
\ee 
we can rewrite Eq.~(\ref{Frim5}) as
\begin{eqnarray}
\label{Frim6}
&&\frac{H^2(z)}{H_0^2}=\left[\Omega_{m,0} \left(1+z\right)^{3}+\Omega_{r,0} \left(1+z\right)^{4}+\Omega_{\Lambda,0}\right]\\[2mm]
\nonumber
&&\times\left\{1+\beta\left[\Omega_{m,0} \left(1+z\right)^{3}+\Omega_{r,0} \left(1+z\right)^{4}+\Omega_{\Lambda,0}\right]\right\}+\mathcal{O}(\beta^2)\,.
\end{eqnarray}
The value of $\Omega_{\Lambda,0}$ 
can be set by imposing the flatness condition $H(0)/H_0=1$, 
which gives
\be
\label{flatness}
\Omega_{\Lambda,0}=1-\Omega_{m,0}-\Omega_{r,0}-\beta+\mathcal{O}(\beta^2)\,.
\ee

A relevant quantity in understanding the dynamics governing the Universe's
evolution is the deceleration parameter
\be
\label{q}
q=-1-\frac{\dot H}{H^2}=-1+\left(1+z\right)\frac{\partial_z H}{H}\,.
\ee
For $q>0$, the expansion of the Universe is decelerating, which implies that
the gravitational pull of matter dominates over any other influences, causing
the expansion to slow down over time. On the other hand, $q<0$ signifies an accelerating Universe. The discovery of the accelerated expansion, supported by
observations of distant type Ia Supernovae~\cite{meas1,meas2} and the CMB~\cite{SDSS:2005xqv}, has led to the acceptance of the existence of some exotic
component that exerts negative pressure and counteracts the gravitational attraction of matter. 

Implications of the zero-point length on the acceleration of the Universe can be studied by plugging Eq.~\eqref{Frim2} into~\eqref{q}. We get
\begin{eqnarray}
\label{q2}
q&=&\frac{\Omega_{m,0}\left(1+z\right)^3+2\Omega_{r,0}\left(1+z\right)^4-2\Omega_{\Lambda,0}}{2\left[\Omega_{m,0}\left(1+z\right)^3+\Omega_{r,0}\left(1+z\right)^4+\Omega_{\Lambda,0}\right]}
\\[2mm]
\nonumber
&&+\,\frac{\beta}{2}\left[3\Omega_{m,0}\left(1+z\right)^3+4\Omega_{r,0}\left(1+z\right)^4\right]+\mathcal{O}(\beta^2)\,.
\end{eqnarray}
At present time, this reduces to
\be
\label{q0}
q_0=-1+\frac{3}{2}\Omega_{m,0}+2\Omega_{r,0}+\beta\left(3\Omega_{m,0}+4\Omega_{r,0}\right)+\mathcal{O}(\beta^2)\,,
\ee
where we have used the condition~\eqref{flatness}. Notice that, 
in the limiting case where $\beta=0$,  we obtain $q_0=-1+\frac{3}{2}\Omega_{m,0}+2\Omega_{r,0}$, which correctly reproduces the $\Lambda$CDM expression. 

It might be interesting to constrain $\beta$ using the observational value 
$q_0=-0.56^{+0.04(0.09)}_{-0.04(0.08)}$~\cite{CapozDag}. Substitution into
Eq.~\eqref{q0} with $\Omega_{m,0}\simeq0.3$ and $\Omega_{r,0}\simeq5\times10^{-5}$
yields $\beta<\mathcal{O}(10^{-1})$. Additionally,  for such range of values of $\beta$, the transition from deceleration to acceleration phase would occur at redshift $z_t\in[0.48,0,75]$, which aligns with the estimate $z_t=0.789^{+0.186}_{-0.165}$ derived from the combined CC+BAO+SNeIa dataset (see also~\cite{Naik:2023ykt}). It should be noted, however, that the above constraint on $\beta$
provides a very weak condition. This could somehow be expected, since
it is inferred from measurements of cosmic parameters at present time. A more stringent bound will be derived below through the study of inflation.

\subsubsection{Matter dominated era}

Let us now investigate the impact of the zero-point length on the dynamics of the scale factor. For the matter dominated era, neglecting the effects of radiation and cosmological constants, Eq.~(\ref{Frim5}) can be approximated as
\begin{equation} \label{Frim7}
\frac{H^2}{H_0^2}=\Omega_{m,0} a^{-3}\left(1+\beta \Omega_{m,0}
a^{-3}\right)+\mathcal{O}(\beta^2).
\end{equation}
The explicit form of the scale factor as a
function of time is obtained by integrating the above relation. We find
\begin{eqnarray} 
\label{am1}
a(t)=\left[\frac{9}{4}\Omega_{m,0}\left(H_0^2t^2-\frac{4}{9}\beta
\right)
\right]^{\frac{1}{3}}\,.
\end{eqnarray}
where we have set the integration constant equal to zero. 
This is indeed the case if we assume $a\rightarrow0$ as $t\rightarrow0$. 

\begin{figure}[t]
\begin{center}
\includegraphics[width=8cm]{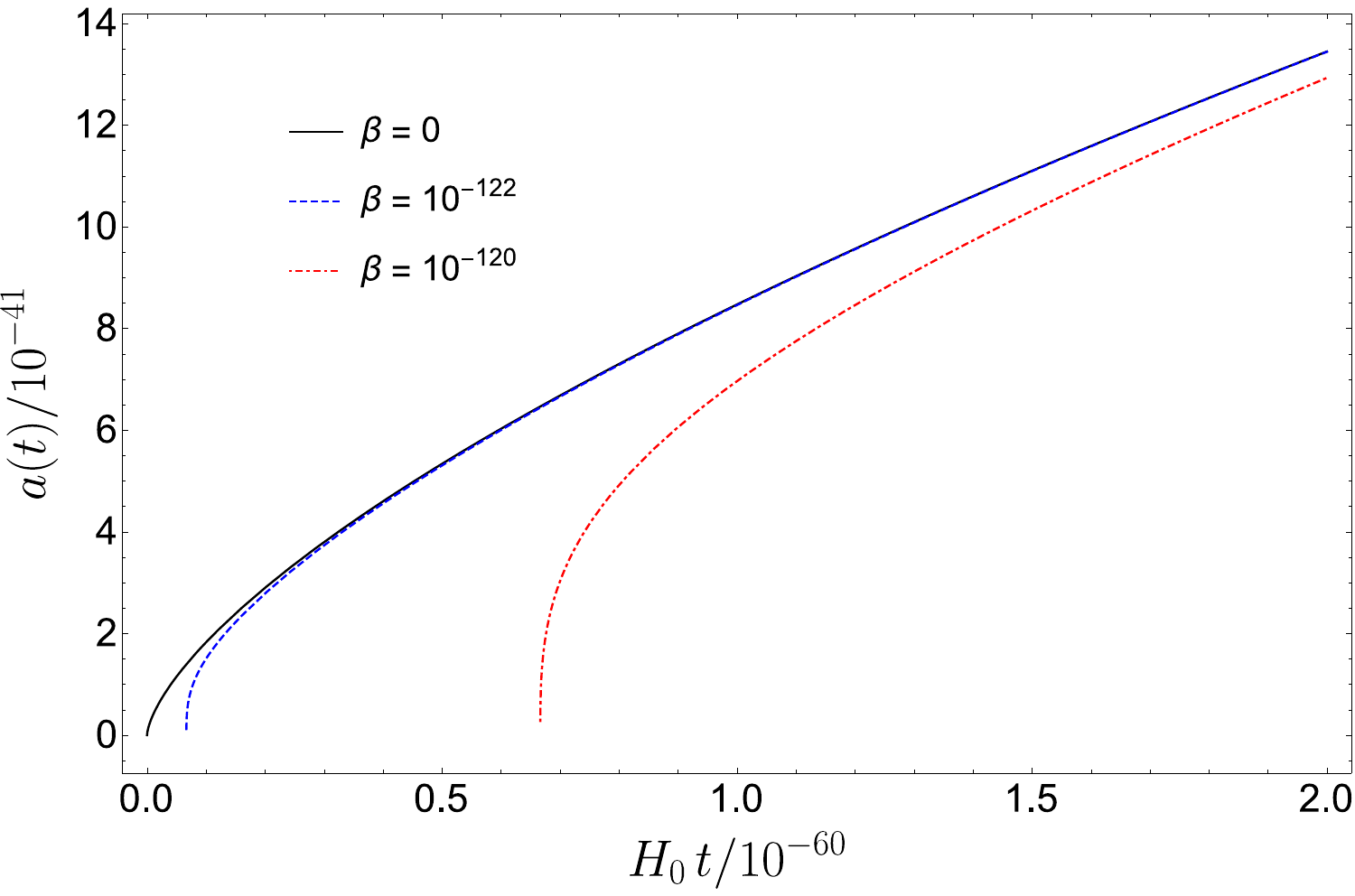}
\caption{Plot of $a(t)$ versus $H_0 t$ for the matter dominated era and various values of $\beta$. We set $\Omega_{m,0}\simeq0.3$.}
\label{am}
\end{center}
\end{figure}

The plot in Fig.~\ref{am} shows that the modified scale factor 
lies below the standard curve for $t$ small enough and
vanishes for $H_0t=2\sqrt{\beta}/3>0$. Below this threshold, 
Eq.~\eqref{am1} apparently takes complex values. This is, however, 
a ``border'' regime, since we are extrapolating our model to scales
comparable or lower than the zero-point length, where the approximation~\eqref{Frie4} fails. Clearly, such phenomenon indicates a breakdown 
of the conventional Cosmology approaching $l_0$ and the need for a comprehensive quantum gravity framework. 
On the other hand, $a(t)$ approaches the standard behavior
as the Universe evolves for sufficiently large $t$, since the zero-point length corrections become increasingly negligible. 
In the limiting case where $\beta=0$, one recovers $a(t)\sim t^{2/3}$,
which is the well-known solution in standard Cosmology for the
matter dominated Universe. 

\subsubsection{Radiation dominated era}

\begin{figure}[t]
\begin{center}
\includegraphics[width=8cm]{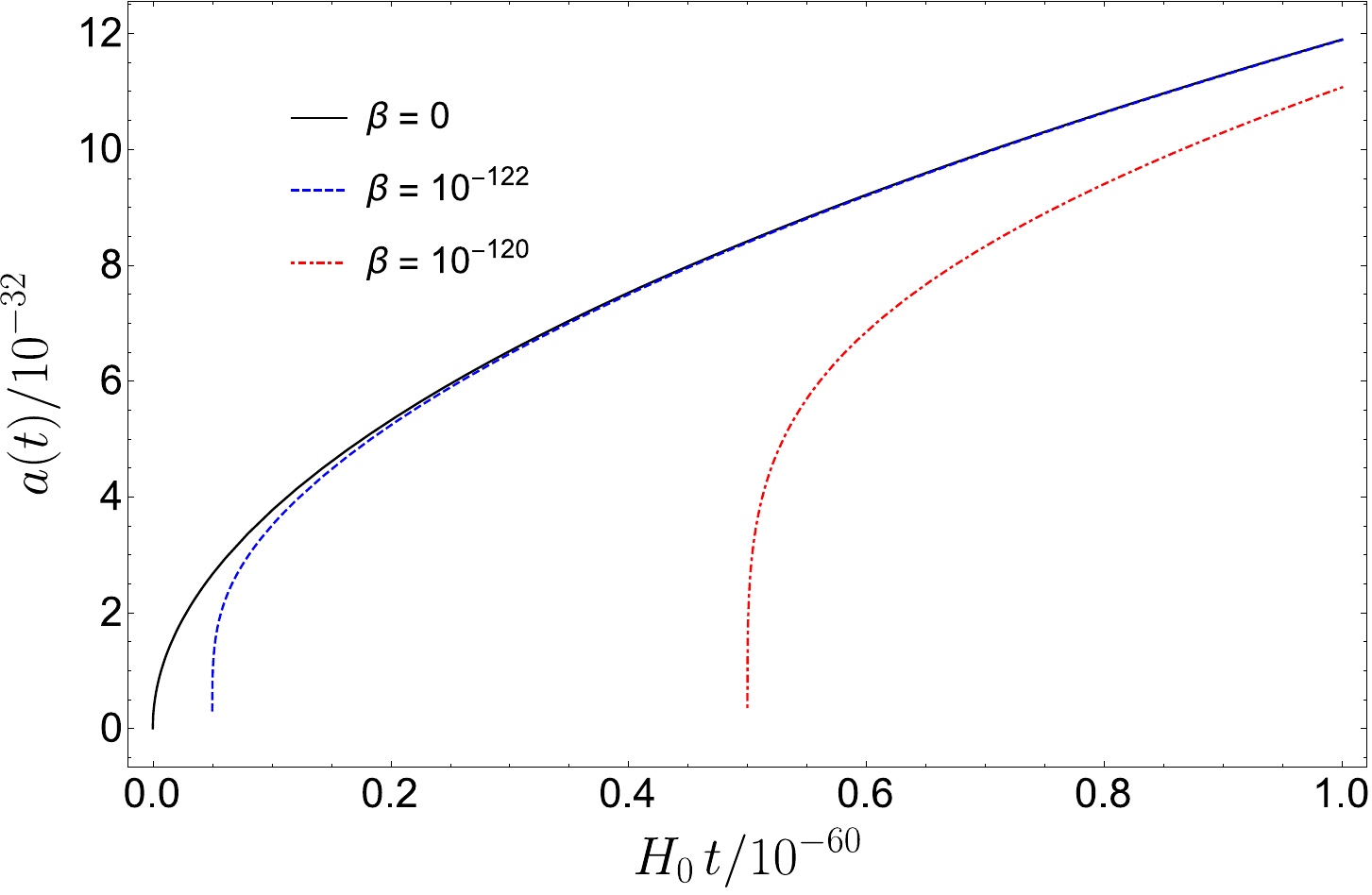}
\caption{Plot of $a(t)$ versus $H_0 t$ for the matter dominated era and various values of $\beta$. We set $\Omega_{r,0}\simeq 5\times 10^{-5}$.}
\label{ar}
\end{center}
\end{figure}

For the radiation dominated era, the Friedmann equation~(\ref{Frim5})
can be rewritten as
\begin{equation} \label{Frim8}
\frac{H^2}{H_0^2}=\Omega_{r,0} a^{-4}\left(1+\beta \Omega_{r,0}
a^{-4}\right)+\mathcal{O}(\beta^2).
\end{equation}
In this case the modified scale factor is obtained as
\begin{equation} \label{ar2}
a(t)=\left[4 \Omega_{r,0} \left(H_0^2 t^2 -\frac{\beta}{4}\right
)\right]^{1/4}\,. 
\end{equation}
where again we have set the integration constant equal to zero. 
The curves in Fig.~\ref{ar} display the same qualitative behavior
as discussed in the previous case, both for small and large $t$. 
When $\beta=0$, one recovers $a(t)\sim t^{1/2}$, which is the
expected result in standard Cosmology for the radiation dominated
epoch.

For later convenience, we notice that the $a$-dependence in Eq.~\eqref{Frim8}
can be converted to a temperature-dependence through the
condition $T a(t)=T_0$, where $T_0\simeq 3\,\mathrm{K}$ is the average temperature of the observable Universe at present time. The latter holds true for any $\beta$, 
as it can be checked by using the relation $\rho\sim T^4$ along with Eqs.~\eqref{rr} and~\eqref{ar2}. In terms of redshift, we can equivalently write
\be
\label{zT}
\frac{T}{T_0}=z+1\,.
\ee

\subsection{Cosmographic Analysis}
The depiction of the Universe, as established by the cosmological principle, is referred to as cosmography~\cite{WeinbergBook}. Cosmography deals with the study of geometric properties, embedding dimensionless higher derivative terms of the scale factor~\cite{Visser}. For the FRW Universe, the latter can be expanded
as 
\be
\label{Taylor}
a(t)=1+\sum_{n=1}^{\infty}\frac{1}{n!}\frac{d^na}{dt^n}\left(t-t_0\right)^n\,.
\ee 
The coefficients of this Taylor expansion are usually referred to as cosmographic parameters. It is easy to check that the first two parameters are related with the Hubble rate and deceleration parameter, respectively. 
On the other hand, the third derivative of the scale factor with respect to cosmic time
gives the jerk parameter~\cite{Jerk1,Jerk2}
\be
\label{jp}
j=\frac{1}{a H^3}\frac{d^3a}{dt^3}=q\left(2q+1\right)+\left(1+z\right)\frac{dq}{dz}\,,
\ee
which is useful in understanding the acceleration of the Universe
and, in particular, the departure of a cosmological model from the $\Lambda$CDM.


\begin{figure}[t]
\begin{center}
\includegraphics[width=8.4cm]{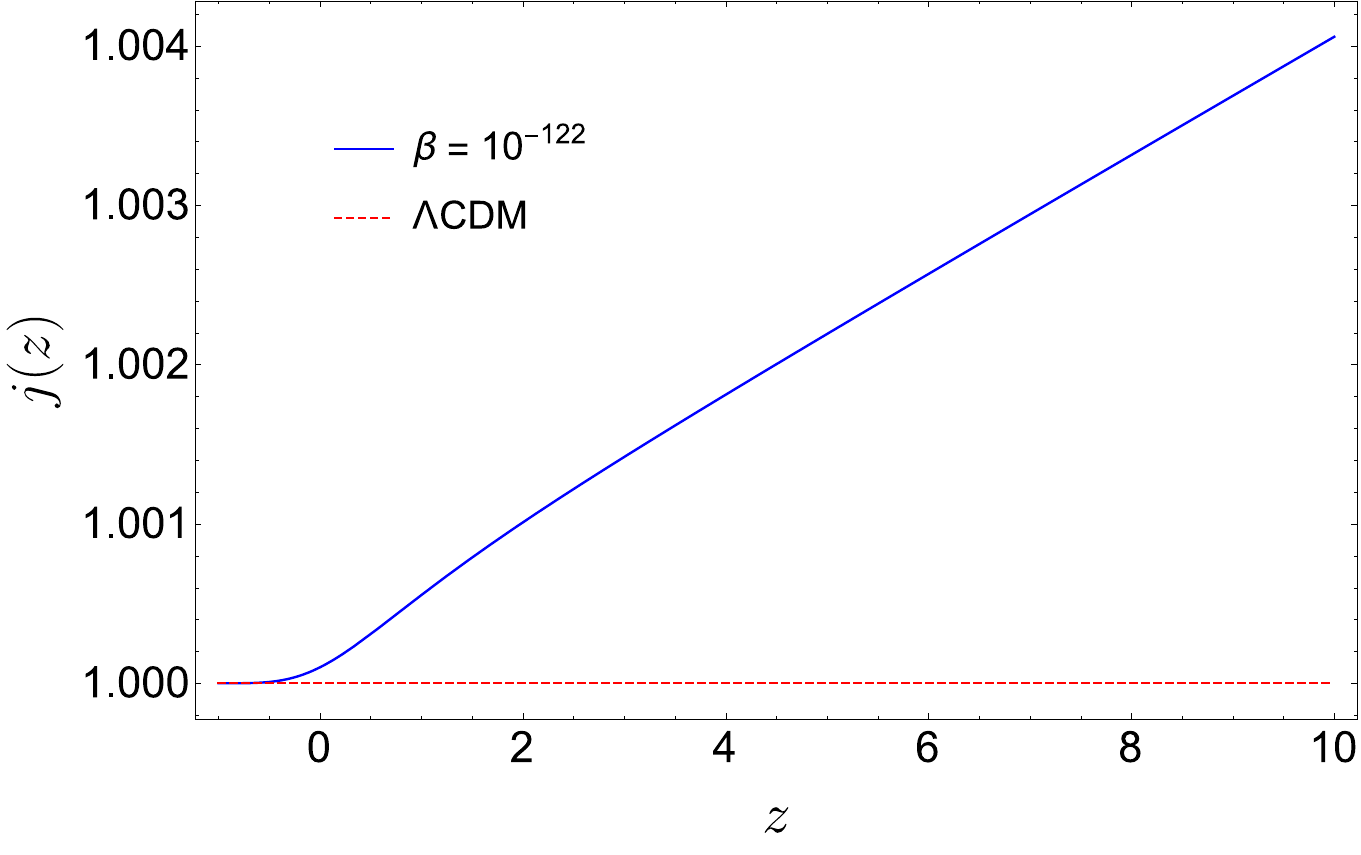}
\caption{Plot of $j(z)$ versus $z$ (blue solid curve). The red dashed line sets the  $\Lambda$CDM value $j=1$. We set $\Omega_{m,0}\simeq0.3$ and $\Omega_{r,0}\simeq 5\times 10^{-5}$.}
\label{jerk}
\end{center}
\end{figure}

The evolution of $j$ as a function of redshift is plotted in Fig.~\ref{jerk} (blue solid curve) in comparison with the $\Lambda$CDM prediction $j=1$ (red dashed line). The analytical computation is easily performed by plugging Eq.~\eqref{q2} into~\eqref{jp}. 
As expected, our model mostly deviates from the $\Lambda$CDM at higher redshift, while the standard description is recovered approaching the present time. This observation prompts the anticipation of an alternative description of early Universe in the presence of zero-point length. Furthermore, we find $j>0$, which is consistent with an accelerated expansion.

Although the first three cosmographic parameters are sufficient to determine the overall kinematics of the Universe~\cite{CapozCosm}, it is sometimes convenient to 
consider extra terms in the expansion~\eqref{Taylor}. For example, 
the snap parameter $s$ is a dimensionless quantity that describes the fourth derivative of the scale factor of the Universe with respect to cosmic time. It provides information about the rate of change of acceleration in the expansion of the Universe,  thus characterizing the dynamics of cosmic acceleration. Following~\cite{Pan:2017ent}, we have
\be
s=\frac{1}{a H^4}\frac{d^4a}{dt^4}=-j\left(2+3q\right)-\left(1+z\right)\frac{dj}{dz}\,.
\ee
Clearly, for the $\Lambda$CDM model $s=-(2+3q)$ since $j=1$. Thus, 
the departure of $ds/dq$ from $-3$ quantifies the deviation of the evolution of the Universe from the $\Lambda$CDM dynamics. 

\begin{figure}[t]
\begin{center}
\includegraphics[width=8.2cm]{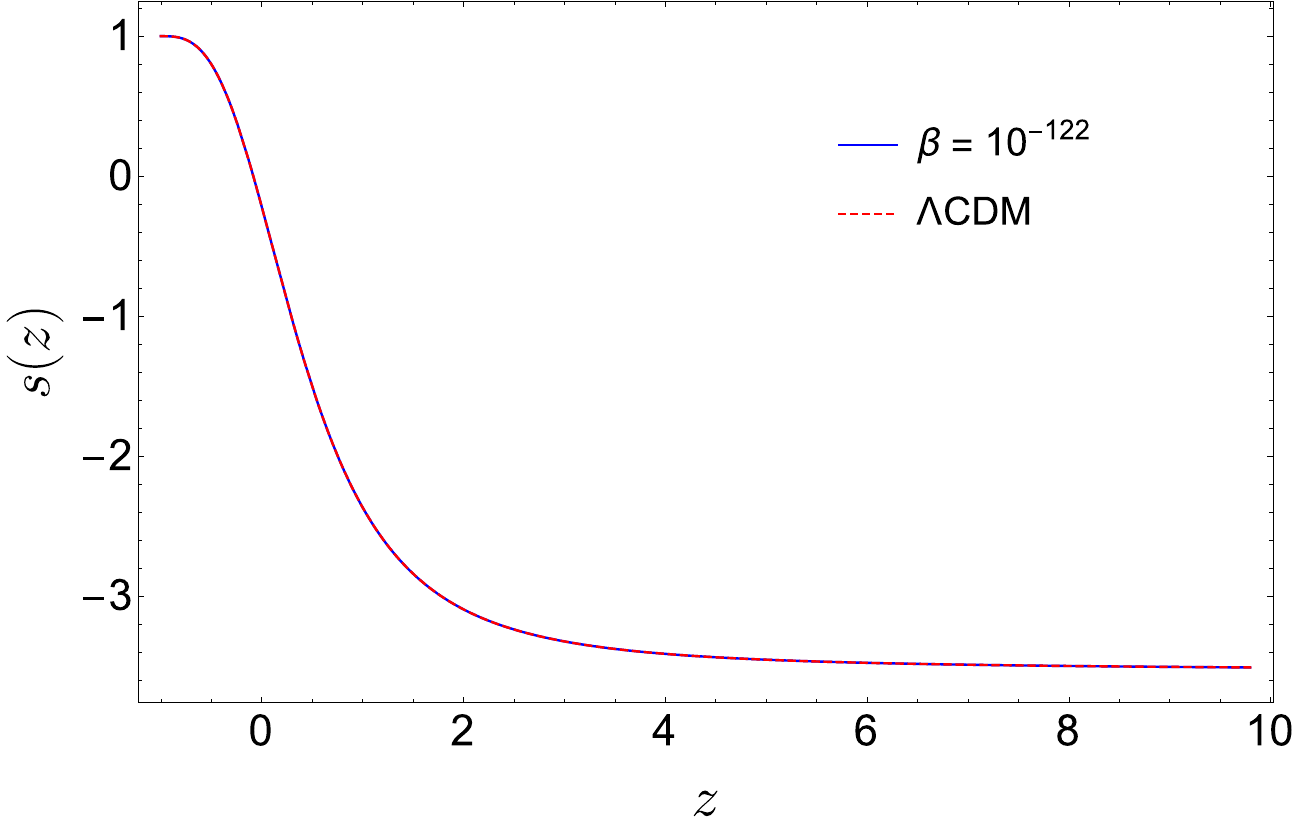}
\caption{Plot of $s(z)$ versus $z$ (blue solid curve). The red dashed curve gives the $\Lambda$CDM prediction. We set $\Omega_{m,0}\simeq0.3$ and $\Omega_{r,0}\simeq 5\times 10^{-5}$.}
\label{snap}
\end{center}
\end{figure}

By using Eqs.~\eqref{q2} and~\eqref{jp}, we infer the evolution of $s$ in the presence of zero-point length (see Fig.~\ref{snap}). For such small values of $\beta$,
deviations from the $\Lambda$CDM model are hardly appreciable in this case even at high redshifts.  Observing the behavior of $s$, it is evident that 
it assumes negative values in the early Universe, transitioning to positive values 
as the evolution progresses. At present time $s_0\simeq-0.2$, consistently with the observational estimate of the deceleration parameter $q_0\simeq-0.6$. 

\begin{figure}[t]
\begin{center}
\includegraphics[width=8.2cm]{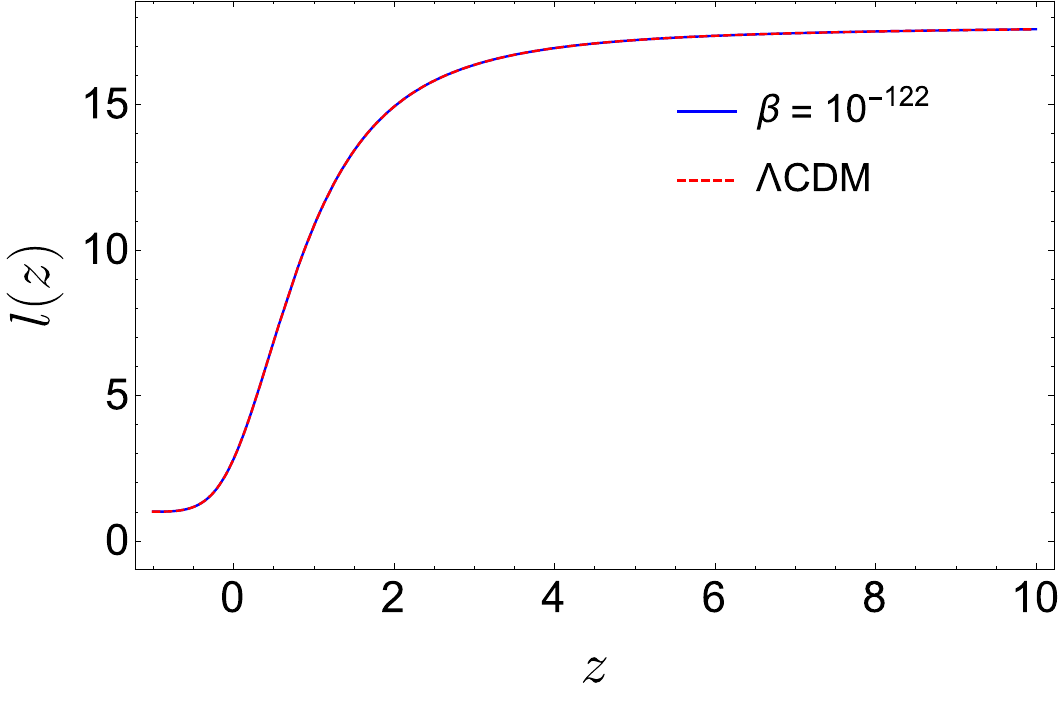}
\caption{Plot of $l(z)$ versus $z$ (blue solid curve). The red dashed curve gives the $\Lambda$CDM prediction. We set $\Omega_{m,0}\simeq0.3$ and $\Omega_{r,0}\simeq 5\times 10^{-5}$.}
\label{lerk}
\end{center}
\end{figure}

Let us finally consider the lerk parameter, which is given by the fifth derivative of the scale factor as~\cite{Pan:2017ent}
\be
l=\frac{1}{a H^5}\frac{d^5a}{dt^5}=-s\left(3+4q\right)-\left(1+z\right)\frac{ds}{dz}\,.
\ee
The evolution of $l$ is plotted in Fig.~\ref{lerk}, 
which displays that the lerk
parameter exclusively maintains positive values. 
Similar to the jerk parameter $j$,
both the snap parameter s and the lerk parameter $l$ tend to unity during
the later stages of the Universe's evolution. 
This trend emphasizes the consistent dynamics exhibited by these parameters in our model.

\section{Slow-roll inflation in zero-point length Cosmology}
\label{inflation}

Since the zero-point length is expected to primarily influence the evolution of the 
early Universe, in this Section we investigate its implications for inflation. 
This phase is usually described in terms of a hypothetical scalar field $\phi$ - the inflaton - which is thought to have been the driving force behind the inflationary expansion. The dynamics of the inflaton field is governed by the (spatially homogenous) potential $V(\phi)$. 

Under the canonical scalar field assumption, the model lagrangian takes the usual form $\mathcal L=X-V(\phi)$, where $X=-\frac{1}{2}h^{\mu\nu}\partial_\mu\phi\hspace{0.2mm}\partial_\nu\phi$ is the kinetic term. The energy density and pressure for the inflation are
\begin{eqnarray}
\label{def1}
\rho_\phi&=&\frac{\dot\phi^2}{2}+V(\phi)\,,\\[2mm]
\label{def2}
p_\phi&=&\frac{\dot\phi^2}{2}-V(\phi)\,.
\end{eqnarray}
The continuity equation~\eqref{Cont} implies the dynamics
\be
\label{KG}
\ddot \phi + 3H \dot\phi + \partial_\phi V(\phi) =0\,,
\ee 
where $\partial_\phi V\equiv\frac{dV}{d\phi}$. 

The successful models of inflation are typically characterized by a very flat potential
along which the inflaton rolls down towards a minimum. If the evolution of the field occurs so slowly that the kinetic energy is negligible compared to the potential term, then the inflaton exhibits a cosmological constant-like behavior, leading to an exponential expansion of the Universe. During this phase, which is referred to as \emph{slow-roll inflation}, matter and radiation components are diluted away rapidly, allowing us to neglect all sources of energy density other than the inflaton. 

Mathematically speaking, the slow-roll approximation is expressed by the conditions~\cite{Luciano:2023roh,Keskin,OdSR}
\begin{eqnarray}
\label{SRC1}
\ddot \phi &\ll& H \dot \phi\,,\\[2mm]
\label{SRC2}
\frac{\dot\phi^2}{2} &\ll& V(\phi)\,.
\end{eqnarray}
The second relation effectively states that $V(\phi)$ is nearly constant, since the inflaton changes very slowly in time. This implies that the Hubble rate remains approximately constant. Therefore, it proves useful to introduce the dimensionless parameter
\be
\label{slow1}
\epsilon \equiv - \frac{\dot H}{H^2}\,,
\ee
where it is understood that $\epsilon\ll1$.
Additionally, to keep track of the relative change of $\epsilon$
per Hubble time, we define a second 
slow-roll parameter as
\be
\label{slow2}
\eta \equiv \frac{\dot \epsilon}{H \epsilon}\,.
\ee
For a sufficiently long inflation, it is likewise needed $\eta\ll1$.

The total amount of expansion that occurred during inflation is quantified
by the number of \emph{e}-folds~\cite{efolds} 
\be
\label{efolds}
\mathcal{N}=\int Hdt=\int \frac{H}{\dot \phi}d\phi\,,
\ee 
from the horizon crossing $\phi_c$ to the end of inflation defined by $\epsilon(\phi_f)=1$. For typical inflationary models, a sensible working hypothesis is $\mathcal{N}$ larger than $\sim 50-60$~\cite{Liddlebis}, which weakly depends on the inflation scale and the thermal history after inflation. However, an extremely long duration of inflation is not excluded, and
appears even necessary in certain scenarios. For example, the stochastic axion framework~\cite{SAM,SAM2}, the relaxation scenario~\cite{Relax} and a quintessence model~\cite{Quint} require $\log_{10}\mathcal{N} \sim\mathcal{O}(10)$, depending on model parameters. Recently, a simple single-field model
has been proposed in the context of stochastic inflation with 
an average $e$-folds $\langle\mathcal{N}\rangle\sim {10^{10}}^{10}$~\cite{StocInf}.  
In the following, we shall consider more traditional and experimentally favored scenarios involving $\mathcal{O}(10^2)\lesssim\mathcal{N}\lesssim\mathcal{O}(10^3)$.


According to the inflationary paradigm, the rapid growth of the Universe
caused fluctuations of the scalar field to be stretched to large scales and, upon crossing the horizon, to freeze in. In the Fourier space, 
such fluctuations are characterized by a power spectrum with spectral index~\cite{Liddle} 
\be
\label{spind}
n_s-1=2\eta-6\epsilon\,,
\ee

There is another kind of density perturbations, the tensor modes, which are fluctuations in the fabric of spacetime. Similarly
to the scalar modes, their spectrum is expected to follow a power-law behavior (see Sec.~\ref{GW} for more details). The ratio of the tensor to scalar spectra 
is given by
\be
\label{ratio}
r=16\epsilon\,.
\ee

The parameters~\eqref{spind},~\eqref{ratio} play a crucial role, as for the possibility 
to constrain or rule out inflationary 
models by comparison with observations. 
In particular, the latest Planck Public Release 4, BICEP/Keck Array 2018, Planck CMB lensing and BAO data set the experimental bounds $r<0.037$~\cite{rconst}
and $n_s\simeq0.97$~\cite{SpecCons1}. 

To study how the zero-point length affects the above parameters, 
we now consider the modified Friedmann equation~\eqref{Frie6}, where we neglect
for simplicity the cosmological constant. Using Eq.~\eqref{def1} for the inflaton energy density along with the first slow-roll condition~\eqref{SRC1}, we obtain
\be
H(\phi)\simeq\sqrt{\frac{8\pi V}{3}}\left(1+\frac{4\pi\beta}{3H_0^2}V\right)+\mathcal{O}(\beta^2)\,,
\ee
where the $\phi$-dependence of $V$ has been omitted in the RHS.

On the other hand, by differentiating Eq.~\eqref{Frie6} with respect to $t$
and using the conditions~\eqref{def1},~\eqref{def2}, we find
\be
\label{HpSF}
\dot H\simeq-4\pi\left(1+\frac{16\pi\beta}{3H_0^2}V\right)\dot\phi^2+\mathcal{O}(\beta^2)\,.
\ee
Notice that the time derivative $\dot\phi$ can be expressed in terms of $\phi$
through Eq.~\eqref{KG}, which in the slow-roll approximation simplifies to 
\begin{eqnarray}
\dot\phi&\simeq&-\frac{1}{3H}\partial_\phi V\\[2mm]
\nonumber
&=&\left(-\frac{1}{2\sqrt{6\pi}}+\sqrt{\frac{2\pi}{27}}\frac{\beta}{H_0^2}V\right)\frac{\partial_\phi V}{\sqrt{V}}\,+\,\mathcal{O}(\beta^2)\,.
\end{eqnarray}

\subsection{Power-law potential}
The latter equation indicates that the dynamics of $\phi$ is ruled by the potential $V(\phi)$. It should be noted that two different approaches are typically considered in studying inflation. The \emph{dynamically motivated} approach involves reconstructing $V$ by requiring the Universe to evolve following a specific dynamics, which is set by a fixed time-dependence of the Hubble rate. From this perspective, inflationary models are classified according to the corresponding Universe evolution. As an example, in~\cite{DMA} the scale factor is set to $a(t)=t^m$ for a suitable choice of the power term $m$. Nevertheless, the disadvantage is that the predicted scalar-to-tensor ratio exceeds the constraint set by the BICEP and Planck data. Moreover, the slow-roll parameters turn out to be constant. On the other hand, the \emph{potential motivated} method involves setting the inflaton potential first and then solving the ensuing Friedmann equations. Such an approach finds wide application in extended theories gravity~\cite{Liddle,Keskin,Luciano:2023roh,NojiOd,DiValentino:2016ziq,Adhikari:2020xcg}, where deviations of the dynamics of the early Universe from GR are admitted. Inflationary models are in this case categorized based on the form of the potential, which is fixed a priori. 

In order to discuss a concrete model of inflation, in the following we consider the 
potential motivated approach. Specifically, we assume the power-law form
\be
\label{pot}
V=V_0 \phi^n\,,
\ee
where the amplitude $V_0$ and the power term $n$ are both positive constants. 
Following~\cite{Liddle,Luciano:2023roh}, $V_0$ is set to unity. Although its actual value may vary in a wide range~\cite{Matarrese}, this assumption is justified by the fact that the magnitude of $V_0$ is expected not to alter the power-law function during inflation. Furthermore, experimental evidence favors inflationary models with $n\sim\mathcal{O}\left(10^{-1}\div1\right)$, while $n\ge2$ tends to be ruled out in the minimal coupling scenario. We shall here consider $n=1$. 

With the ansatz~\eqref{pot}, the dimensionless parameter~\eqref{slow1} is found to be
\be
\label{eps}
\epsilon(\phi)\simeq \frac{1}{16\pi\phi^2}+\mathcal{O}(\beta^2)\,,
\ee
which also allows us to derive the inflaton at the end of inflation as
\be
\epsilon(\phi_f)=1 \,\,\Longrightarrow\,\, \phi_f\simeq\frac{1}{4\sqrt{\pi}}+\mathcal{O}(\beta^2)\,.
\ee

The spectral index $n_s$ and the tensor-to-scalar ratio $r$ are typically
evaluated at the horizon crossing, where the fluctuations of the inflaton freeze and remain nearly constant~\cite{Liddle,Keskin,Luciano:2023roh,LucSheLamb}. From Eq.~\eqref{efolds}, the inflaton at the horizon crossing is obtained as
\be
\label{fieldhc}
\phi_c\simeq\frac{1}{4}\sqrt{\frac{f(\mathcal{N})}{\pi}}+\frac{\beta}{18H_0^2}\left[\frac{1-\sqrt{f^3(\mathcal{N})}}{\sqrt{f(\mathcal{N})}}\right]+\mathcal{O}(\beta^2)\,,
\ee
where we have used the shorthand notation 
\be
f(\mathcal{N})\equiv1+4\mathcal{N}\,.
\ee
Therefore, the slow-roll parameters~\eqref{spind},~\eqref{ratio} become
\begin{eqnarray}
\nonumber
\hspace{-4mm}r_\beta&=&16\epsilon(\phi_c)\\[2mm]
\label{rbetaeq}
&\simeq&\frac{16}{f(\mathcal{N})}+\frac{64\sqrt{\pi}\beta}{9H_0^2}\,\frac{\sqrt{f^3(\mathcal{N})}-1}{f^2(\mathcal{N})}+\mathcal{O}(\beta^2)\,,\\[4mm]
\nonumber
\hspace{-4mm}n_s{_\beta}&=&1+2\eta(\phi_c)-6\epsilon(\phi_c)\\[2mm]
&\simeq&1+\frac{2}{f(\mathcal{N})}-\frac{8\sqrt{\pi}\beta}{9H_0^2}\,\frac{5\sqrt{f^3(\mathcal{N})}+1}{f^2(\mathcal{N})}+\mathcal{O}(\beta^2)\,.
\label{nbetaeq}
\end{eqnarray}

\begin{figure}[t]
\begin{center}
\includegraphics[width=8.4cm]{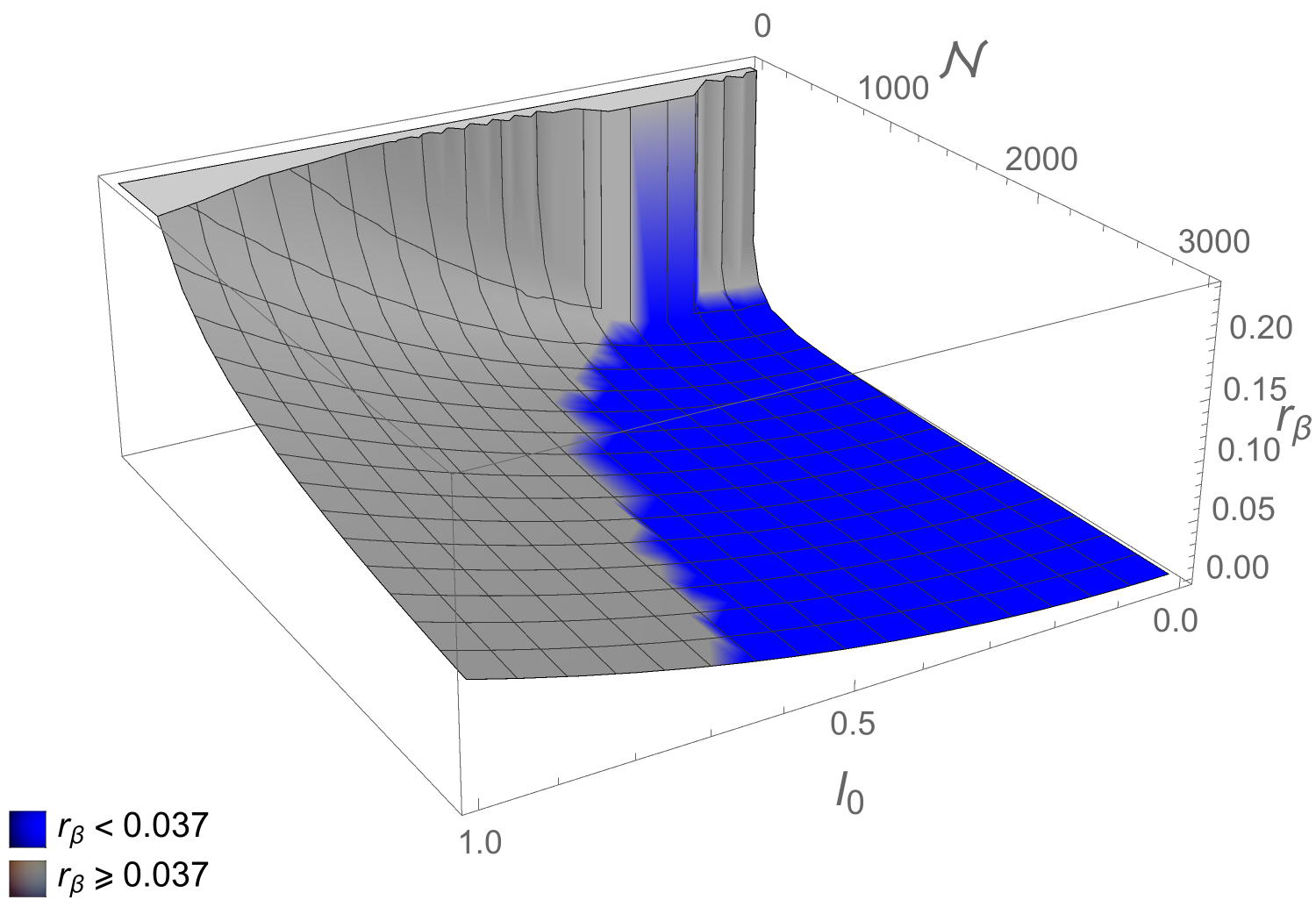}
\caption{3D Plot of $r_\beta$ versus $\mathcal{N}$ and $l_0$. The grey region is excluded by observational constraints~\cite{rconst}.}
\label{rbeta}
\end{center}
\end{figure}

In order to constrain the parameter $\beta$ (or, equivalently, the zero-point length $l_0$), let us compare the above expressions with experimental bounds on $n_s$ and $r$ (see below Eq.~\eqref{ratio}). Assuming the expected-inflation duration $\mathcal{N}\sim\mathcal{O}(10^3)$, it can be inferred from Fig.~\ref{rbeta}  that 
\be
\label{boundrb}
r_\beta<0.037 \Longrightarrow \l_0\lesssim\mathcal{O}(1)\,,
\ee
corresponding to $\beta\lesssim\mathcal{O}(10^{-122})$. 
This turns into the more stringent condition $l_0\lesssim\mathcal{O}(10^{-1})$ $(\beta\lesssim\mathcal{O}(10^{-124}))$ for a total number of $e$-folds $\mathcal{N}\sim\mathcal{O}(10^2)$. Direct substitution into Eq.~\eqref{nbetaeq} shows that these bounds are consistent with the observational measurement $n_s\simeq0.97$ too. 

Interestingly enough, the above constraints
align with the original assumption $l_0\lesssim1$ of the zero-point length from quantum fluctuations of the background spacetime~\cite{Padm} and the recent result~\cite{td1}  in quantum corrected black holes from String T-Duality. 

\section{Structure formation}
\label{pert}
A key challenge in modern Cosmology lies in comprehending the origin and dynamics of perturbation growth during the early Universe, as these fluctuations eventually give rise to the formation of the observed large-scale structures. It is widely accepted that these structures emerge from the amplification of tiny initial density fluctuations through gravitational instability during cosmic evolution. These fluctuations gradually grow until they become sufficiently robust to break away from the background expansion and collapse into gravitationally-bound systems (see~\cite{Mukhanov:1990me} and~\cite{Malik:2008im} for a recent review). 

An effective framework for investigating the growth of matter perturbations and structure formation in the linear regime is the Top-Hat Spherical Collapse (SC) model~\cite{Abramo:2007iu}. This approach considers a uniform and spherically symmetric perturbation in an expanding background, describing the growth of perturbations within a spherical region using the Friedmann equations of gravity~\cite{Planelles:2014zaa}. Employing this model, 
in the following we study the impact of the zero-point length Cosmology on 
the growth of cosmic perturbations and the power spectrum of scalar fluctuations  at the early stages of the Universe.

\subsection{Growth of perturbations in linear regime}
As a first step, by using the relation
\be
\frac{\ddot a}{a}=\dot H + H^2\,,
\ee
let us recast the second modified Friedmann equation
as
\be
\label{add}
\frac{\ddot a}{a}=-\frac{4\pi}{3}\rho\left(1+\frac{32\pi\beta}{3H_0^2}\rho\right)\,.
\ee
To apply the Top-Hat SC model, we focus on a 
spherical region of radius $a_p$ and with a homogenous density $\rho_c$. 
Assume that, at time $t$, $\rho_c(t)=\rho(t)+\delta\rho$, where $\delta\rho$
is the density fluctuation. The conservation equation in such simplified spherical region is given by~\cite{Abramo:2007iu}
\be
\dot\rho_c+3h\rho_c=0\,,
\ee
where $h=\dot a_p/a_p$  denotes the local expansion rate inside the spherical 
region. The second Friedmann equation applied to this region reads
\be
\frac{\ddot a_p}{a_p}=-\frac{4\pi}{3}\rho_c\left(1+\frac{32\pi\beta}{3H_0^2}\rho_c\right)\,.
\ee

It is now useful to define the density contrast of the fluid by the relation
\be
\label{densco}
\delta\equiv\frac{\rho_c}{\rho}-1=\frac{\delta\rho}{\rho}\,.
\ee
Since the fluctuation $\delta\rho$ is expected to be smaller than $\rho$, 
we can assume $\delta\ll1$ (linear regime). This is a reasonable approximation, as the initial stages of the gravitational collapse are well described 
by the linear regime for all but the very smallest sales~\cite{Abramo:2007iu}. 

Differentiating Eq.~\eqref{densco} with respect to $t$, we find
\be
\dot\delta=3\left(1+\delta\right)\left(H-h\right),
\ee
where we have used the continuity equation. Further
differentiation gives
\be
\label{dynper}
\ddot\delta=3\left(1+\delta\right)(\dot H-\dot h)+\frac{\dot\delta^2}{1+\delta}\,.
\ee

Equation~\eqref{dynper} governs the dynamics of matter
perturbations in the Top-Hat SC model. 
It can be further manipulated by noting that
\be
\dot H-\dot h=h^2-H^2+\frac{4\pi\rho}{3}\left(1+\frac{64\pi\beta}{3H_0^2}\rho\right)\delta+\mathcal{O}(\delta^2)\,,
\ee
where we have neglected higher order terms in $\delta$.  
After some algebra, substitution into Eq.~\eqref{dynper} gives 
\be
\label{dyn}
\ddot\delta+2H\dot\delta-{4\pi\rho}\left(1+\frac{64\pi\beta}{3H_0^2}\rho\right)\delta+\mathcal{O}(\delta^2)=0\,.
\ee

Changing the time variable to the sale factor $a$,  we have
\begin{eqnarray}
\dot\delta&=&aH\delta'\,,\\[2mm]
\ddot\delta&=&a^2H^2\delta''+aH\left(H+aH'\right)\delta'\,,
\end{eqnarray}
where the prime denotes derivative with respect to $a$. 
Upon substituting into Eq.~\eqref{dyn}, we obtain 
\be
\label{dynbis}
\delta''+\delta'\left(\frac{3}{2a}-\frac{4\pi\beta}{H_0^2}\frac{\rho}{a}\right)-\delta\left(\frac{3}{2a^2}+\frac{28\pi\beta}{H_0^2}\frac{\rho}{a^2}\right)+\mathcal{O}(\delta^2)=0.
\ee
It is easy to check that the standard dynamics is recovered for $\beta\rightarrow0$~\cite{Abramo:2007iu}. 

\begin{figure}[t]
\begin{center}
\includegraphics[width=8.2cm]{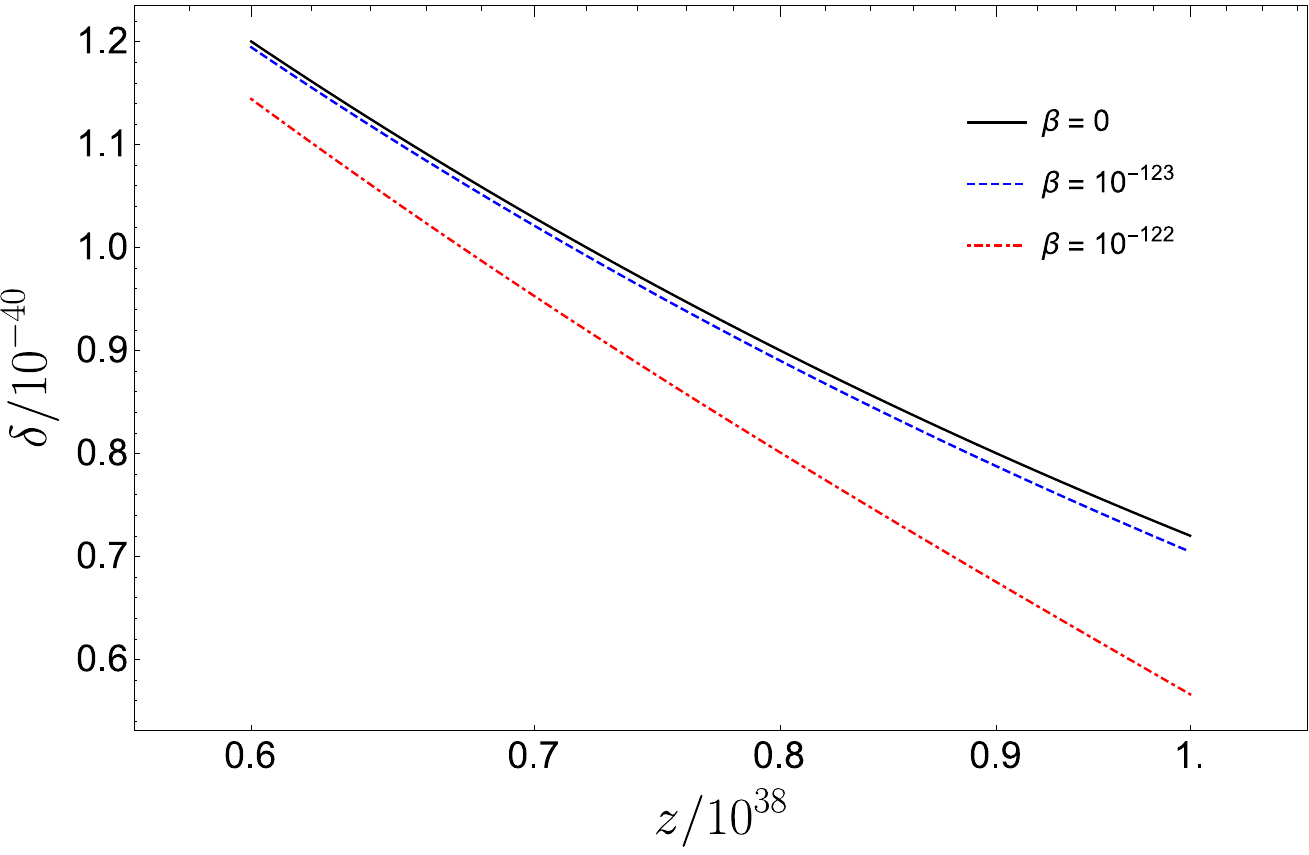}
\caption{Log-linear plot of $\delta$ versus $z$, for various values of $\beta$. 
We set the same initial conditions as in~\cite{Abramo:2007iu}. }
\label{gro}
\end{center}
\end{figure}

Equation~\eqref{dynbis} has been solved numerically. The solution is 
plotted in Fig.~\ref{gro} for various values of $\beta$ consistent with Eq.~\eqref{boundrb}.  As expected, effects of the zero-point length mainly manifest as $z$ increases, with the matter density contrast growing more slowly than the standard behavior (black solid curve) in the very early Universe. 
Therefore, unlike the classical framework, where
small-scale fluctuations evolve unhindered, such fluctuations are smoothed out
due to the quantum gravity corrections associated with the zero-point length. This smoothing effect limits the growth of fluctuations, leading to a slower increase in the density contrast. A similar suppression has been discussed in~\cite{Galaxies}, assuming a running gravitational constant. Other studies appear in~\cite{Oth1,Oth2,Oth3} in the context of full quantum and/or extended gravity. 

It should be noted that the above conclusions provide preliminary, but still insightful results toward understanding the effects of zero-point length on the evolution of the density contrast and structure formation. Indeed, for $l_0$ around the Planck length, 
deviations from the classical model could become appreciable at very high redshift, where the linear regime may break down and higher order corrections should be taken into account. A more refined analysis going beyond the linear regime
is reserved for the future. 

Before moving on, it is interesting to observe that the evolution of 
matter overdensity $\delta$ is closely related to the $\sigma_8$ tension, 
which quantifies the gravitational matter clustering from the amplitude of the linearly evolved power spectrum at the scale of  $8h^{-1}\,\mathrm{Mpc}$. 
This tension arises from the fact that the Cosmic Microwave Background
(CMB) estimation~\cite{Planck:2018vyg} differs from the SDSS/BOSS direct measurements~\cite{eBOSS:2018cab,BOSS:2016wmc}. The later growth 
of $\delta$ in Fig.~\ref{gro} for non-vanishing $\beta$ indicates 
a less efficient clustering of matter over the time in the presence of 
a minimal length, which provides a possible mechanism to alleviate the $\sigma_8$ tension~\cite{Heis}. A similar outcome in the framework of modified gravity 
has been found in~\cite{sig8} using a Tsallis-like deformation of the Bekenstein-Hawking entropy. 

\subsection{Power spectrum}
We now look into the primordial power spectrum for scalar perturbations, 
which provides an indirect probe of inflation or other structure-formation mechanisms. The simplest inflationary models predict almost purely adiabatic primordial perturbations with a nearly scale-invariant power spectrum given by~\cite{Har,Zel}
\be
\label{HZ}
P_{\zeta}(k) = A_S\left(\frac{k}{k_*}\right)^{n_s-1}\,,
\ee
where $A_s\simeq10^{-9}$ is the power amplitude and $k_*$
a characteristic pivot scale. 

The form~\eqref{HZ} is the statistically preferred by Planck, since it yields a good fit to data with a remarkably small number of parameters. However,
it does not provide the best observational fit. Improvements can be obtained by allowing for a deviation from the simple power spectrum,   
either considering a running spectral index or different functional forms~\cite{deBlas}.
This is mainly due to the possibility of suppressing the primordial power spectrum at large scales. 

A paradigmatic modification considered by the Planck Collaboration consists 
in a broken-power-law (bpl) spectrum of the form~\cite{deBlas}
\be
\label{mpmod}
P^{bpl}_\zeta\propto
       \left(\frac{k}{k_*}\right)^{n_s^{bpl}-1+\delta}\,, 
  \ee
for $k$ below a certain cutoff scale. The best fit is obtained for 
$\delta=1.14$~\cite{deBlas}. 

Interestingly enough, the zero-point length model 
naturally fits the profile~\eqref{mpmod}. To check this, 
let us recast Eq.~\eqref{nbetaeq} as
\be
\label{nsbis}
n_s{_\beta}=n_s + g_\beta\,,
\ee
where $n_s$ is the usual spectral index in standard Cosmology ($\beta=0$)
and 
\be
g_\beta\equiv -\frac{8\sqrt{\pi}\beta}{9H_0^2}\,\frac{5\sqrt{f^3(\mathcal{N})}+1}{f^2(\mathcal{N})}+\mathcal{O}(\beta^2)
\ee
is the zero-point length correction. 
The latter term has the same role as the $\delta$-shift 
in the bpl spectrum~\eqref{mpmod}. Said another way, 
the correction~\eqref{mpmod} is reproduced in the zero-point length Cosmology
while still maintaining the simple power spectrum~\eqref{HZ}. Indeed,
upon replacing the modified spectral index~\eqref{nsbis},  
the exponent of Eq.~\eqref{HZ} becomes
\be
n_s-1\,\,\longrightarrow\,\, n_s{_\beta}-g_\beta-1\,.
\ee
The RHS fits Eq.~\eqref{mpmod}, provided that 
\be
\label{fitdelta}
n_s{_\beta}-g_\beta-1=n_s^{bpl}-1+\delta\,.
\ee
Setting $n_s{_\beta}$ and $n_s^{bpl}$ equal to the 
experimental value~\cite{SpecCons1} and using Eq.~\eqref{beta}, 
we find
\be
l_0=\frac{f(\mathcal{N})}{\pi^{1/4}}\sqrt{\frac{3\delta}{2+10f^{3/2}(\mathcal{N})}}\,,
\ee
which, for $\mathcal{O}(10^2)\lesssim N\lesssim\mathcal{O}(10^3)$ and $\delta=1.14$,  gives $l_0\sim\mathcal{O}(1)$, in compliance with Eq.~\eqref{boundrb}. 

Therefore, zero-point length Cosmology stands out as 
an alternative characterization for the bpl spectrum, which is still consistent with latest observations. 

\section{Primordial Gravitational Waves}
\label{GW}
The first ever detection of gravitational waves (GWs) by the LIGO~\cite{LIGOScientific:2014pky} and VIRGO~\cite{VIRGO:2014yos} collaborations has opened up fresh avenues for investigating the Universe. 
From a theoretical perspective, GWs originate from three main sources categorized based on their generation mechanisms: astrophysical, cosmological and inflationary sources. Astrophysical GWs are notoriously influenced by the mass of the emitting object, with massive compact objects like black holes expected to produce strong GWs. For the minimum mass $M_\odot$, for instance, the maximal frequency would be around $10\, \mathrm{kHz}$. On the other hand, various cosmological phenomena in the early Universe, such as first-order phase transitions at $10^{-2}\,\mathrm{GeV}$ or just below the Grand Unified Theory scale, can produce GWs with frequency around $10^{-5}\,\mathrm{Hz}$ or roughly lower than the $\mathrm{GHz}$ range, respectively. Finally, primordial GWs generated during inflation span frequencies from $10^{-18} \,\mathrm{Hz}$ to $1\,\mathrm{GHz}$ (see~\cite{Ito:2022rxn} for more details).  While current observatories are primarily focused on astrophysical signals, it is clear that the exploration of high-frequency GWs represents a promising yet challenging test bench for theories beyond the standard Cosmology, that could not be tested otherwise. Starting from these premises, in what follows we compute the spectrum of primordial gravitational waves (PGWs) in the zero-point length Cosmology and compare it with predictions of the standard scenario. 

\subsection{Standard Cosmology}

GWs in the linearized approximation are described as metric perturbations on a curved background.  Using the transverse traceless conditions, the equation of motion for tensor perturbation at the first order takes the form~\cite{Watanabe:2006qe}
\be
\label{hdyn}
\ddot h_{ij} +  3H\dot h_{ij} -  \frac{\nabla^2}{a^2}h_{ij} =  16\pi \Pi_{ij}^{TT}\,,
\ee
where $\Pi_{ij}^{TT}$ is the transverse-traceless
part of the anisotropic stress tensor
\be
\Pi_{ij} =  \frac{T_{ij}-p\hspace{0.3mm} g_{ij}}{a^2}\,,
\label{TPE}
\ee
and $T_{ij}, g_{ij}$ and $p$ denote the stress-energy tensor, metric tensor and  background pressure, respectively (latin indices run over the three spatial coordinates of spacetime). 

We now focus on GW signals with frequencies higher than $10^{-11}\,\mathrm{Hz}$~\cite{Weinberg:2003ur}. To solve Eq.~\eqref{hdyn}, it proves convenient to work in the Fourier space, where
\be
\label{hij}
h_{ij}(t,\vec{x}) =  \sum_{\lambda}\int\frac{d^3k}{\left(2\pi\right)^3}\hspace{0.2mm}h^\lambda(t,\vec{k})\hspace{0.2mm}\epsilon^\lambda_{ij}(\vec{k})\hspace{0.2mm}e^{i\vec{k}\cdot\vec{x}}\,.
\ee
The two independent polarizations have been denoted by $\lambda=+,\times$, while
$\epsilon^\lambda$ is the (normalized) spin-2
polarization tensor obeying $\sum_{ij}\epsilon^\lambda_{ij}\epsilon^{\lambda'*}_{ij}=2\delta^{\lambda\lambda'}$ (it is understood that arrows in Eq.~\eqref{hij} denote three-vectors). In this way, the tensor perturbation $h^\lambda(t,\vec{k})$ can be factorized as
\be
h^\lambda(t,\vec{k}) =  h_{prim}^\lambda(\vec{k})X(t,k)\,,
\ee
where $X(t,k)$ and $h_{prim}^\lambda$ denote the transfer function and 
amplitude of the primordial tensor perturbations, respectively. 

In this setting, the tensor power spectrum is~\cite{Bernal:2020ywq}
\be
\mathcal{P}_T(k) =  \frac{k^3}{\pi^2}\sum_\lambda\Big|h^\lambda_{prim}(\vec k)\Big|^2  =  \frac{2}{\pi^2}\hspace{0.3mm}G\hspace{0.3mm} H^2\Big|_{k=aH}\,,
\ee
while the dynamics~\eqref{hdyn} takes the form of a damped harmonic
oscillator-like equation
\be
X'' \ + \ 2\hspace{0.2mm}\frac{a'}{a}X' \ + \ k^2X \ = \ 0\,. 
\ee
With abuse of notation, the prime here indicates derivative with respect to the conformal time $d\tau=dt/a$.

\begin{figure}[t]
\begin{center}
\includegraphics[width=8.5cm]{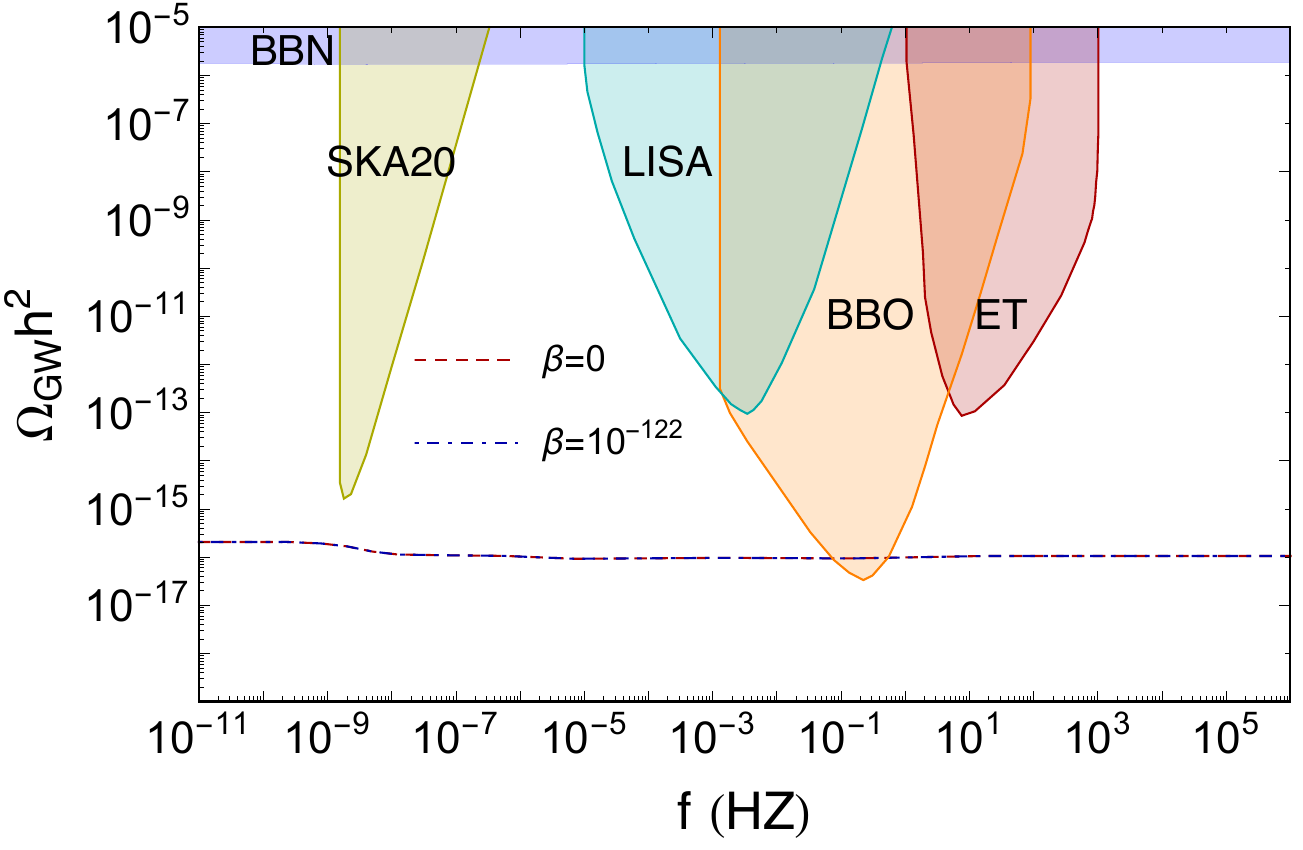}
\caption{Plot of the PGW spectrum versus the frequency $f$, for  $n_T=0$ and $A_S\sim10^{-9}$, consistently with the Planck observational constraints at the CMB scale~\cite{ConPlanck}. The colored regions delineate the projected sensitivities for several GW observatories~\cite{Breitbach:2018ddu}.}
\label{GWplot2}
\end{center}
\end{figure}

In standard Cosmology, the relic density of PGW from first-order tensor perturbation
reads~\cite{Bernal:2020ywq,Watanabe:2006qe}
\begin{eqnarray}
\nonumber
\Omega_{GW}
(\tau,k)&=&\frac{[X'(\tau,k)]^2}{12a^2(\tau)H^2(\tau)}\,\mathcal{P}_T(k)\\[2mm]
&\simeq&\left[\frac{a_{hc}}{a(\tau)}\right]^4\left[\frac{H_{hc}}{H(\tau)}\right]^2\frac{P_T(k)}{24}\,,
\label{Ttps}
\end{eqnarray}
where we have averaged over oscillation periods in the second step, i.e.
\be
X'(\tau,k) \simeq  k\hspace{0.2mm} X(\tau,k) \simeq  \frac{k\hspace{0.3mm} a_{hc}}{\sqrt{2}a(\tau)} \simeq 
\frac{a^2_{hc}\hspace{0.3mm}H_{hc}}{\sqrt{2}a(\tau)}\,.
\ee
Moreover, we have used 
\be
\label{k}
k=2\pi f=a_{hc}H(a_{hc})
\ee 
for the horizon crossing moment\footnote{In~\cite{Guzzetti:2016mkm} 
the PGW spectrum is parametrized in terms of the reheating temperature, revealing 
the occurrence of a knee between $10^2\,\mathrm{Hz}$ and $10^6\,\mathrm{Hz}$  for $10^{10}\,\mathrm{GeV}\lesssim T_R\lesssim10^{14}\,\mathrm{GeV}$. Assuming the energy scale associated with the horizon crossing during inflation lies between $10^{16}\,\mathrm{GeV}$ and $10^{19}\,\mathrm{GeV}$~\cite{ConPlanck}, in the present study this change in slope is expected above $10^6\,\mathrm{Hz}$. For our next purposes, however, it is enough
to restrict to the low-frequency range $10^{-11}\,\mathrm{Hz}\lesssim f\lesssim 10^{6}\,\mathrm{Hz}$, where the PGW spectrum in GR remains approximately flat (see Fig.~\ref{GWplot2}).}. Therefore,  the PGW relic density at present time is  
\be
\Omega_{GW}(\tau_0,k)h^2\simeq\left[\frac{g_*(T_{\mathrm{hc}})}{2}\right]\left[\frac{g_{*s}(T_0)}{g_{*s}(T_{\mathrm{hc}})}\right]^{4/3}\frac{\mathcal{P}_T(k)\Omega_{r}(T_0)h^2}{24}\,,\,,
\label{Spt}
\ee
where $h$ is the dimensionless (or reduced) Hubble constant, while $g_*(T)$ and $g_{*s}(T)$ denote the effective numbers of relativistic degrees of freedom that contribute to the radiation energy
density $\rho$ and entropy density $s$, respectively, i.e.
\be
\rho_r=\frac{\pi^2}{30}g_*(T)T^4\,,\quad\,\, s_r=\frac{2\pi^2}{45}g_{*s}(T)T^3\,.
\ee

Next, the scale dependence of the tensor power spectrum is given by
\be
\mathcal{P}_T(k)=A_T\left(\frac{k}{k_*}\right)^{n_T}\,,
\ee
where the amplitudes of the tensor and scalar perturbations are related via the 
the tensor-to-scalar ratio $r$ as $A_T=r A_S$, while
$n_T$ is the tensor spectral index. 

The plot in Fig.~\ref{GWplot2} displays the behavior of the PGW spectrum in Eq.~\eqref{Spt} (red dashed line) versus the frequency (we have restored correct units for comparison with literature).  The colored regions delineate the
projected sensitivities for the Square Kilometre Array (SKA) telescope~\cite{Janssen:2014dka}, LISA interferometer~\cite{LISA:2017pwj}, Einstein Telescope (ET) detector~\cite{Sathyaprakash:2012jk} and the successor Big Bang Observer (BBO)~\cite{Crowder:2005nr}. Moreover, the Big Bang Nucleosynthesis (BBN) bound is set  by the constraint on the effective number of neutrinos~\cite{Boyle:2007zx,Stewart:2007fu}.

\subsection{Zero-point length Cosmology}

We now study implications of the zero-point length for the PGW spectrum. 
Toward this end, we make use of the modified Friedmann equation~\eqref{Frim5}
and the relation~\eqref{k}. Following~\cite{Bernal:2020ywq}, we can write from Eq.~\eqref{Ttps}
\begin{eqnarray}
\nonumber
\Omega_{\mathrm{GW}}(\tau,k)&\hspace{-1mm}\simeq\hspace{-1mm}& \left[\frac{a_{hc}}{a(\tau)}\right]^4\left[\frac{H_{hc}}{H_{\mathrm{GR}}(\tau)}\right]^2\left[\frac{H_{\mathrm{GR}}(\tau)}{H(\tau)}\right]^2\frac{P_T(k)}{24}\\[2mm]
 \nonumber
&&\hspace{-20mm}=\,\Omega^{\mathrm{GR}}_{\mathrm{GW}}(\tau,k)\left[\frac{H_{\mathrm{GR}}(\tau)}{H(\tau)}\right]^2\left[ \frac{a_{hc}}{a_{hc}^{GR}}\right]^4\left[ \frac{a^{GR}(\tau)}{a(\tau)}\right]^4\left[ \frac{H_{hc}}{H_{hc}^{GR}}\right]^2   \,, \\
\label{eq:PGWBar0}
\end{eqnarray}
with
\be
k \ = \ a_{\mathrm{hc}}^{\mathrm{GR}}H^{\mathrm{GR}}(a_{\mathrm{hc}}^{\mathrm{GR}})\,.
\ee
Here, standard quantities in General Relativity have been denoted by GR, e.g., 
$\Omega^{\mathrm{GR}}_{\mathrm{GW}}(\tau,k)$ is the PGW relic density

\begin{figure}[t]
\begin{center}
\includegraphics[width=8.5cm]{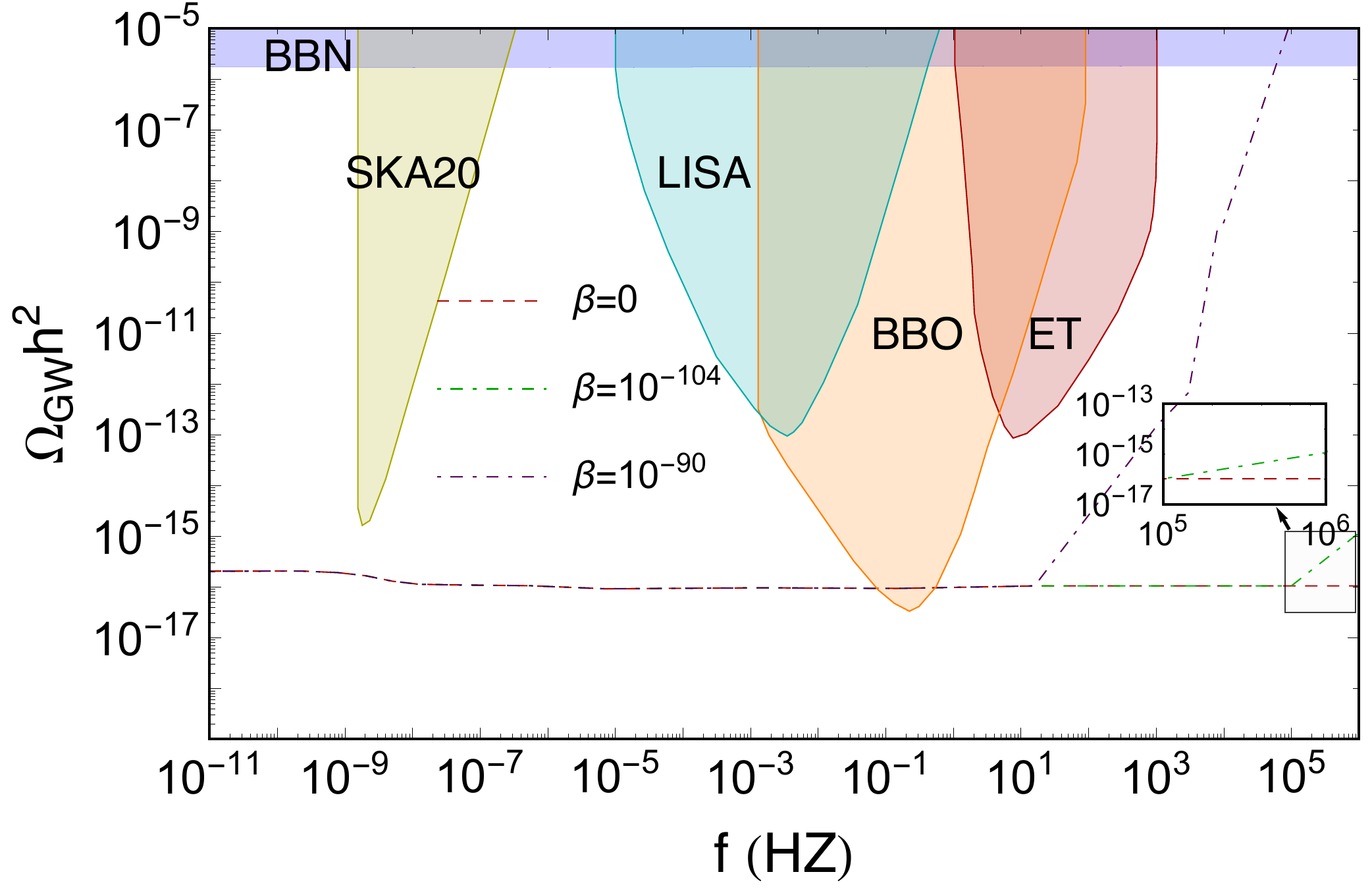}
\caption{Plot of the PGW spectrum versus the frequency $f$ for various values of $\beta$, $n_T=0$ and $A_S\sim10^{-9}$, consistently with the Planck observational constraints at the CMB scale~\cite{ConPlanck}. The colored regions delineate the projected sensitivities for several GW observatories~\cite{Breitbach:2018ddu}.}
\label{GW3}
\end{center}
\end{figure}

Next, using the flatness condition $H(z)/H_0=1$, we obtain
\begin{equation}
 \Omega_{\mathrm{GW}}(\tau_0,k)
\ \simeq \ \Omega^{\mathrm{GR}}_{\mathrm{GW}}(\tau_0,k)\left[ \frac{a_{hc}}{a_{hc}^{GR}}\right]^4\left[ \frac{H_{hc}}{H_{hc}^{GR}}\right]^2   \,. 
\label{eq:PGWBar} 
\end{equation}
The modified PGW spectrum for $\beta\sim\mathcal{O}(10^{-122})$ (blue dot-dashed line) is plotted in Fig.~\ref{GWplot2}. Unfortunately,   if the zero-point length is around the Planck scale, no significant deviation from GR (red dashed curve) could be detected in upcoming GW observatories, at least up to frequencies around $10^{-3}\,\mathrm{GHz}$. 
Nevertheless, it is interesting to note that appreciable effects might still be observed in the frequency range $f\gtrsim10^{-4}\,\mathrm{GHz}$, provided that 
$\beta\gtrsim10^{-104}$ ($l_0\gtrsim10^9$) (green dot-dashed curve in Fig.~\ref{GW3}). If, on the one hand, this range of $\beta$ values does not fit the constraint set above through the inflation measurements, on the other hand it could be understood by allowing for a running zero-point length parameter. 
Although not contemplated in the original model, 
such a scale-dependent behavior would be fully justified, as it is consistent with the typical case occurring in quantum field theory and quantum gravity under renormalization group applications, where all parameters and coupling constants are dynamical in tandem with the energy scale.  
Moreover, the coupling constant in conventional quantum gravity decreases with increasing distance, in such a way that the larger the distance scale, the smaller the 
coupling, and vice-versa (cf. e.g.~\cite{Hooft}).
Notice that a similar framework
has been investigated in~\cite{Petroz} assuming a fluctuating minimal length in the Generalized Uncertainty Principle, and in~\cite{Runde} by setting a varying exponent in Barrow holographic dark energy. 
Clearly, such a scenario may open up the possibility of new physics to be tested through future measurements of high-frequency GWs above the LIGO/VIRGO range. This analysis goes beyond the scope of the present work and will be explored elsewhere.

Moreover, a zero-point length exceeding $l_0\simeq10^{16}$ ($\beta\simeq10^{-90}$, purple dot-dashed curve in Fig.~\ref{GW3}) appears disfavored (although not yet excluded) in our model, since it would entail a dramatic departure from GR for even GWs below $10\,\mathrm{KHz}$.
This prediction is somehow consistent with the experimental evidence from particle physics~\cite{LHC} (we remind that subatomic distances have been directly measured
up to $10^{-18}\,\mathrm{m}$ and beyond, corresponding to length scales  $l\simeq10^{17}$ in our units). The same rapid increase of the PGW spectrum at low-frequencies has been exhibited in extended scalar-tensor and extradimensional gravity~\cite{Bernal:2020ywq}.

\section{Discussion and Conclusions \label{Clos}}

An approach to overcome the singularity problem in black hole
physics is using string inspired T-duality model. In this method
one can remove the singularity at the center of black hole by
taking into account a zero-point length in the gravitational
potential, which come from momentum space propagator induced by
the path integral duality \cite{td1}. Based on this, in the present paper we have explored the effects of the zero-point length in
the cosmological setup. We first derived the
modified Friedmann equations due to zero-point length by applying
the first law of thermodynamics on the apparent horizon. Then, we have
explored the cosmological consequences of this model and disclosed
the influences of the zero-point length on the history of the
Universe in both the matter and radiation dominated era. 
The next effort has been devoted to slow-roll solutions for power-law inflation. 
Comparison of the slow-roll parameters with the observational values has given the constraint $l_0\lesssim\mathcal{O}(10^{-1}\div 1)$, depending on the expected-inflation duration. This is consistent with the original assumption~\cite{Padm} 
and the recent result in quantum corrected black holes from String T-Duality~\cite{td1}. 

Using the Top-Hat Spherical Collapse model, 
we have explored the impact on the growth of matter fluctuations in the linear regime. In spite of this minimal assumption, 
it has been shown that the zero-point length affects the evolution of the density contrast in a non-trivial way, leading to a slower growth of the primordial seed fluctuations into cosmic structures. In this context, we have also analyzed 
implications for the power spectrum of scalar perturbations.
Although the simplest (and statistically preferred) inflationary models predict almost purely adiabatic perturbations with a power-law spectrum, a better observational fit is obtained by either assuming a running spectral index or different functional forms~\cite{deBlas}. We have found that our model naturally
fits the broken-power-law~\eqref{mpmod}, where the role of 
the $\delta$-shift is played by the zero-point length correction (see Eq.~\eqref{fitdelta}). 

We have finally studied the spectrum of primordial gravitational waves
in the early Universe. Although Planck scale corrections to GR are
hardly detectable in the near future, non-trivial effects 
might still be observed at frequencies accessible to the next generation GW experiments if $l_0\gtrsim10^9$. Such a scenario is not that speculative, as it could be understood within the context of a dynamical, energy-scale-dependent behavior for the zero-point length. This would align with the typical case observed
in quantum field theory and quantum gravity under renormalization group considerations  (see, e.g.~\cite{Hooft})..

Further aspects are yet to be investigated. First, we would like to relax our
leading order assumption~\eqref{Frie4} and  possibly develop exact analytical computations. This is essential for a thorough application of our model to the Planck era, which could provide hints toward the formulation of a modified Cosmology in full quantum gravity. Additionally, it would be suggestive to adapt the analysis of Sec.~\ref{inflation} to different inflationary models. A paradigmatic case is the Starobinsky inflation (and related generalizations)~\cite{Staro}, which is largely studied, for instance, for the theory of the origin of the Universe from a quantum multiverse~\cite{Multiv}. 

On the other hand, recent advancements in GW experiments offer a fresh perspective on the nature of gravity in the early Universe. By probing different cosmic scales, these experiments provide a window into potential modifications to Einstein's theory of gravity. For instance, once we detect PGWs, we will gain valuable insights that could help narrow down the parameter space for possible modifications to GR during the early Universe. While the present effort provides a preliminary analysis in this area, it is crucial to emphasize that a detailed comparison of various laboratory/astrophysical gravity tests with PGW detection appears essential. Work is in progress in this direction and will be presented elsewhere.

The era between the end of inflation and the onset of radiation domination in the Universe remains largely unexplored, and there are several UV-complete scenarios that suggest nonstandard Cosmology or modifications to gravity. In this sense, the timing of our study is very pertinent, as the next generation of GW observatories will cover several decades of frequency ranges of various GW amplitudes, allowing for 
the possibility to uncover new physics related to the pre-BBN era.

\acknowledgements
GGL would like to thank L.~Mastrototaro for useful discussion.
He acknowledges the Spanish ``Ministerio de Universidades''
for the awarded Maria Zambrano fellowship and funding received
from the European Union - NextGenerationEU. The
work of AS is supported by Iran National Science Foundation (INSF)
under grant No. 4022705.

\end{document}